\documentclass[12pt]{article}

\usepackage{amsfonts,amssymb,amsmath,amscd}
\usepackage[ps,dvips,matrix,arrow,frame,import,curve,color]{xy}
\newcommand{\ra}{\rightarrow}

\newcommand{\Tr}{{\rm Tr}}

\newcommand{\ZZ}{{\mathbb Z}}
\newcommand{\RR}{{\mathbb R}}
\newcommand{\CC}{{\mathbb C}}

\newcommand{\tQ}{{\tilde{Q}}}

\newcommand{\cL}{{\mathcal L}}
\newcommand{\cO}{{\mathcal O}}

\renewcommand{\part}{\partial}

\newcommand{\bpartial}{{\bar\partial}}

\newcommand{\bpsi}{{\bar\psi}}
\newcommand{\bchi}{{\bar\chi}}
\newcommand{\bxi}{{\bar\xi}}
\newcommand{\blambda}{{\bar\lambda}}

\newcommand{\bz}{{\bar z}}
\newcommand{\bw}{{\bar w}}
\newcommand{\bD}{{\bar D}}

\newcommand{\cV}{{\mathcal V}}
\newcommand{\ot}{\otimes}
\newcommand{\eps}{\epsilon}

\newcommand{\tq}{{\tilde q}}
\newcommand{\barD}{{\bar D}}
\newcommand{\cA}{{\mathcal A}}
\newcommand{\cF}{{\mathcal F}}
\newcommand{\cN}{{\mathcal N}}
\newcommand{\MH}{{{\mathcal M}_H}}
\newcommand{\cE}{{\mathcal E}}
\newcommand{\MV}{{{\mathcal M}_V}}
\newcommand{\frq}{{\mathfrak q}}

\renewcommand{\Re}{{\rm Re}}
\renewcommand{\Im}{{\rm Im}}
\newcommand{\Ka}{{K\"ahler}}
\renewcommand{\a}{{\alpha}}
\renewcommand{\b}{{\beta}}
\newcommand{\cD}{{\mathcal D}}
\newcommand{\cR}{{\mathcal R}}
\newcommand{\ch}{{\rm ch}}
\newcommand{\ad}{{\rm ad}}
\newcommand{\cS}{{\mathcal S}}
\newcommand{\tB}{{\tilde B}}
\newcommand{\cW}{{\mathcal W}}
\newcommand{\g}{{\mathfrak g}}
\newcommand{\LG}{{{}^LG}}
\newcommand{\PP}{{\mathbb P}}
\newcommand{\Ltau}{{{}^L\tau}}
\newcommand{\cM}{{\mathcal M}}
\newcommand{\Det}{{\rm Det}}
\newcommand{\Bun}{{\rm Bun}}
\newcommand{\cC}{{\mathcal C}}
\newcommand{\nn}{\nonumber}
\newcommand{\btau}{{\bar\tau}}

\title{Holomorphic reduction of $\cN=2$ gauge theories, Wilson-'t Hooft operators, and S-duality}

\author{Anton Kapustin\\{\small \it California Institute of Technology, Pasadena, CA 91125,
U.S.A.}}

\begin{document}

\begin{titlepage}

\maketitle

\begin{abstract}
We study twisted $\cN=2$ superconformal gauge theory on a product of
two Riemann surfaces $\Sigma$ and $C$. The twisted theory is
topological along $C$ and holomorphic along $\Sigma$ and does not
depend on the gauge coupling or theta-angle. Upon Kaluza-Klein
reduction along $\Sigma$, it becomes equivalent to a topological
B-model on $C$ whose target is the moduli space $\MV$ of nonabelian
vortex equations on $\Sigma$. The $\cN=2$ S-duality conjecture
implies that the duality group acts by autoequivalences on the
derived category of $\MV$. This statement can be regarded as an
$\cN=2$ counterpart of the geometric Langlands duality. We show that
the twisted theory admits Wilson-'t Hooft loop operators labelled by
both electric and magnetic weights. Correlators of these loop
operators depend holomorphically on coordinates and are independent
of the gauge coupling. Thus the twisted theory provides a convenient
framework for studying the Operator Product Expansion of general
Wilson-'t Hooft loop operators.

\end{abstract}

\vspace{-7in}
\parbox{\linewidth}
{\small\hfill \shortstack{CALT-68-2623}} \vspace{6in}

\end{titlepage}

\section{Introduction}

The observation that many supersymmetric field theories can be
twisted into topological field theories was first made by E. Witten
\cite{WittenTFT} and has proved very fruitful for understanding
properties of such theories as well as their relationship with
mathematics (see e.g.
\cite{Wittentopsigma,Wittenmirror,Vafawitten,WittenSW}). Recently a
twisted version of the $\cN=4$ gauge theory in four dimensions has
been used to provide a physical derivation of the central statements
of the Geometric Langlands Program \cite{KW}. The main goal of this
paper is to explore to what extent the considerations of \cite{KW}
can be generalized to gauge theories with $\cN=2$ supersymmetry.

One obvious motivation is finding analogs of the Geometric Langlands
Program. It is known that Montonen-Olive duality, which implies the
geometric Langlands duality, has close analogs for certain finite
$\cN=2$ gauge theories. The simplest example of such a model is the
so-called Seiberg-Witten theory \cite{SW2} which is $\cN=2$ gauge
theory with gauge group $SU(2)$ and four hypermultiplets in the
fundamental representation. It is natural to inquire whether $\cN=2$
S-duality implies something similar to the geometric Langlands
duality.

A somewhat different motivation stems from the observation made in
\cite{KW} that the twisted $\cN=4$ gauge theory considered there has
topological Wilson-'t Hooft loop observables. These observables are
the most basic observables in any gauge theory, and the twisted
theory, being independent of the gauge coupling, offers an
opportunity to compute some of their properties exactly. For
example, it was shown in \cite{KW} that the OPE algebra of 't Hooft
operators in the theory with gauge group $G$ reproduces the fusion
rules for irreducible representations of the Langlands-dual group
$\LG$. This was achived by identifying the 't Hooft operators with
Hecke operators studied by mathematicians and by exploiting existing
mathematical results on the algebra of Hecke operators (``the
geometric Satake correspondence'' \cite{Ginz,MV1,MV2}). {}From the
physical viewpoint, this is a new nontrivial test of the
Montonen-Olive duality conjecture.

In the untwisted theory, Wilson-'t Hooft operators are labelled
\cite{K} by pairs $(\mu,\nu)\in \Lambda_{cw}\times\Lambda_w$, where
$\Lambda_{cw}$ and $\Lambda_w$ are coweight and weight lattices of
$G$; more precisely, they are labelled by equivalence classes of
such pairs under the action of the Weyl group. The usual Wilson loop
operators correspond to pairs of the form $(0,\nu)$, while 't Hooft
operators have $\nu=0$ and $\mu$ arbitrary. A natural problem is to
compute the algebra of general Wilson-'t Hooft operators with both
$\mu$ and $\nu$ nonzero. Unfortunately, it is not possible to study
this question in the framework of the GL-twisted $\cN=4$ theory
considered in \cite{KW}, for the following reason. The GL-twisted
theory depends on an extra complex parameter $t\in\CC\cup
\{\infty\}$, and for each value of $t$ the weight $\nu$ must be
proportional to the coweight $\mu$, with the coefficient depending
on $t$ and the complexified gauge coupling of the theory. (Here we
assume that the Cartan subalgebra and its dual have been identified
using the Killing metric.) Thus for each particular value of $t$ one
can study only a subset of Wilson-'t Hooft operators.

We will see that there exists a nontopological twist which allows
arbitrary values for $\mu$ and $\nu$ in a given theory. The twisted
theory is still independent of the gauge coupling, so semiclassical
computations of the OPE algebra should give an exact result. We will
also see that such a twist makes sense for an arbitrary finite
$\cN=2$ gauge theory, and this allows one to formulate an analog of
the geometric Langlands duality. More precisely, what can be
generalized to the $\cN=2$ case is the `` classical limit'' of the
usual geometric Langlands duality.

The twist we are going to consider works only for manifolds whose
holonomy group is reduced to $U(1)\times U(1)$. That is, the
four-manifold is a product of two Riemann surfaces $C$ and $\Sigma$.
The twisted theory depends on a parameter $t\in \CC\cup \{\infty\}$
similar to the one in the GL-twisted $\cN=4$ theory. If $t\neq
0,\infty$, the theory is topological on $C$ and holomorphic on
$\Sigma$. That is, it is invariant under arbitrary diffeomorphisms
of $C$, but depends on the complex structure of $\Sigma$. Under the
Kaluza-Klein reduction along $C$ it becomes a chiral CFT on
$\Sigma$. Thus we can attach to any finite $\cN=2$ gauge theory a
chiral algebra. In fact, this construction works in greater
generality: as first noted in \cite{Johansen} and discussed below,
it is sufficient to assume that the theory one starts with has
$\cN=1$ supersymmetry and nonanomalous $U(1)_R$ symmetry.

But in order to have Wilson-'t Hooft operators in the twisted theory
and be able formulate an analog of the geometric Langlands duality,
it is better to start with an $\cN=2$ theory and reduce along
$\Sigma$. The resulting effective field theory on $C$ is a 2d TFT,
namely the B-model whose target is the moduli space of so-called
vortex equations on $\Sigma$. Vortex equations generalize Hitchin
equations to the case when the Higgs field is in the representation
other than the adjoint. Many considerations of \cite{KW} have direct
analogs for such a B-model; in particular 't Hooft operators are
identified with Hecke operators, and they act by functors on the
category of branes on the moduli space of vortex equations.

The values $t=0$ and $t=\infty$ are special in that the twisted
theory does not admit either branes or Wilson-'t Hooft loop
observables of any kind. But such a theory is interesting in its own
right because it is a holomorphic field theory on $C\times\Sigma$.
That is, it depends on the complex structures of both $C$ and
$\Sigma$ and correlators can have holomorphic dependence on
coordinates. (It also depends on the complexified gauge coupling
$\tau$, but only on the nonperturbative level.) Such a theory can be
viewed as a higher-dimensional analog of chiral algebra. It has both
local and nonlocal observables; while the OPE of local observables
is nonsingular due to Hartogs' theorem, this is not so if one also
considers nonlocal observables. {}From the mathematical viewpoint, the
nonlocal observables take values in the sheaf cohomology of certain
holomorphic line bundles on $C\times\Sigma$. Thus twisted
supersymmetric theories suggest a physically motivated definition of
a chiral algebra in two complex dimensions.

Topological reduction of $\cN=2$ gauge theories on a Riemann surface
has been considered for the first time by Bershadsky, Johansen,
Sadov, and Vafa \cite{BJSV}. It was noted there that the reduced
theory is a sigma-model whose target is the moduli space of
nonabelian vortex equations. The main difference between the present
work and \cite{BJSV} is that Bershadsky et al. consider a
topologically twisted theory in four dimensions, while we consider a
nontopological (holomorphic) twist. As a consequence, their
effective sigma-model is an A-model for the moduli space of vortex
equations, while we end up with a B-model with the same target
space.

The organization of the paper is as follows. After a very brief
review of $\cN=2$ gauge theories, we define in section \ref{twists}
a twist of a finite $\cN=2$ theory on $C\times\Sigma$ and study its
properties. Then in section \ref{KK} we perform Kaluza-Klein
reduction along both $C$ and $\Sigma$ and reinterpret our results in
two-dimensional terms. We also briefly comment on holomorphic twists
of $\cN=1$ gauge theories. In section \ref{observables} we discuss
observables in the twisted $\cN=2$ theory: local observables and
their descendants, as well as Wilson-'t Hooft loop observables. In
section \ref{branes} we discuss how Wilson and 't Hooft operators
act on branes in the twisted theory; the general case of mixed
Wilson-'t Hooft operators will be studied elsewhere \cite{future}.
Finally, in section \ref{Sduality} we propose an $\cN=2$ analog of
the Geometric Langlands Program in the special case of the
Seiberg-Witten theory.

We mostly follow the conventions of \cite{KW}. One notable
difference is that we define the covariant derivative to be $D=d+iA$
instead of $d+A$, so that the connection 1-form $A$ is Hermitian.
This is in line with most of the physics literature. Generators
$T^a$ of a compact Lie group $G$ are also taken to be Hermitian, so
that ${\rm Tr } T^a T^b$ is a positive definite metric on the Lie
algebra of $G$. This accounts for some extra minus signs compared to
\cite{KW}. Also, the roles of $C$ and $\Sigma$ are in some sense
reversed compared to \cite{KW}.

\section{Partial twists of finite $\cN=2$ gauge
theories}\label{twists}

\subsection{$\cN=2$ gauge theories}

In this paper we consider $\cN=2$ gauge theory with gauge group $G$
and hypermultiplets in the representation $R$ of $G$. The gauge
group must be compact and semisimple; the representation $R$ is
allowed to be reducible. An irreducible component of $R$ is called a
``flavor''; for example, if $R$ is a sum of $N_f$ copies of an
irreducible representation $R_0$, one says that the theory has $N_f$
flavors of hypermultiplets in representation $R_0$. If we use $N=1$
superfield notation, an $\cN=2$ hypermultiplet in representation $R$
is described by a pair of chiral superfields $\tQ$ and $Q$ in
representations $R$ and and its dual $R^\vee$ respectively. An
$\cN=2$ gauge multiplet is described by an $N=1$ real superfield $V$
and a chiral superfield $\Phi$, both in the adjoint representation
of $G$. The part of the action describing the gauge fields is
$$
I_{gauge}=\frac{1}{2\pi}\Im\left(\tau \int d^4x\, d^2\theta\, \Tr\,
W^\alpha W_\alpha\right)+\frac{\Im\tau}{4\pi}\int d^4x\, d^4\theta\,
\Tr\left(\Phi^\dag e^{2V}\Phi\right).
$$
Here $W_\alpha=-\frac{1}{8}\barD^2 e^{-2V}D_\alpha e^{2V}$, as
usual, and
$$
\tau=\frac{\theta}{2\pi}+\frac{4\pi i}{e^2}
$$
is the complexified gauge coupling which combines the usual gauge
coupling $e$ and the theta-angle. The part of the action
describing the matter fields is
$$
I_{matter}=\int d^4x\, d^4\theta\, \left(Q^\dag e^{2V} Q+\tQ e^{-2V}
\tQ^\dag\right)+\sqrt 2\Re\int d^4x\, d^2\theta\, \tQ \Phi Q.
$$
Here and in what follows we set the hypermultiplet mass terms to
zero.

When we consider the twisted theory, it will be convenient to
rescale all scalar fields (i.e. the lowest components of $\Phi,Q, $
and $\tQ$) by a factor $\sqrt 2$. This is also the convention
adopted in \cite{KW}.

The classical theory has $SU(2)_R\times U(1)_N$ group of
R-symmetries as well as a global symmetry $U(1)_B$ which multiplies
$Q$ and $\tQ$ by opposite phases. In the quantum theory, the
$U(1)_N$ subgroup, under which $\Phi$ has charge $2$ and $Q$ and
$\tQ$ have charge zero, is generically anomalous. The anomaly
cancelation condition is
\begin{equation}\label{ac}
C(G)-C(R)=0,
\end{equation}
where $C(R)$, the index of representation $R$, is defined by
$$\Tr_R T^a T^b=C(R) \delta^{ab},$$
and $C(G)$ is the index of the adjoint representation. The condition
for vanishing of the beta-function is the same as the $U(1)_N$
anomaly cancelation condition. The theories satisfying (\ref{ac})
are superconformal field theories; they are also known as finite
$\cN=2$ theories, since the fields and the gauge coupling do not
require infinite renormalization.

The simplest way to satisfy (\ref{ac}) is to take $R$ to be the
adjoint representation; this gives a theory with $\cN=4$
supersymmetry. For $G=SU(N)$ one can take $R$ to be the sum of $2N$
copies of the fundamental representation; this theory is known as
$\cN=2$ super-QCD (with $2N$ flavors) and has been extensively
studied beginning with the work of N. Seiberg and E. Witten
\cite{SW2} in the case $N_c=2$ and P. Argyres and A. Faraggi
\cite{AF} in general.

In the superconformal case, the gauge coupling $e^2$ does not run
and the theory depends on a single parameter $\tau$ taking values in
the complex upper half-plane. The upper half-plane is acted upon by
the group $SL(2,\RR)$. In the case of $\cN=4$ super-Yang-Mills, the
S-duality conjecture states that the theory is invariant with
respect to an action of a certain discrete subgroup of $SL(2,\RR)$
which is commensurate with $SL(2,\ZZ)$
\cite{MO,Osborn,Giveonetal,Doreyetal,AKS} (if $G$ is simply-laced,
this subgroup coincides with $SL(2,\ZZ)$). The $\cN=4$ S-duality
conjecture follows from the S-duality of type IIB string theory, as
well as from the properties of the $(2,0)$ superconformal field
theory in six dimensions \cite{vafageomorigin}.

What about other superconformal $\cN=2$ theories? In the case of
$\cN=2$ super-QCD with gauge group $SU(N)$ and $2N$ flavors , there
is an S-duality conjecture very similar to the $\cN=4$ S-duality
conjecture~\cite{SW2,Argyres}. In the special case $N=2$, it can be
deduced from the S-duality of Type IIB string theory. The duality
group is again $SL(2,\ZZ)$ in this case. For $N>2$ the conjectured
duality group is the subgroup $\Gamma_0(2)$ of
$SL(2,\ZZ)$~\cite{Argyres}. It is tempting to conjecture that
S-duality exists for all superconformal $\cN=2$ gauge theories.

\subsection{Partial twist along $C$}

Consider a superconformal $\cN=2$ field theory on a Euclidean
four-manifold of the form $M=C\times \Sigma$, where $C$ and $\Sigma$
are Riemann surfaces. The curvature of $M$ in general breaks all
supersymmetry, and if we wish to preserve some of it, we have to
twist the gauge theory so that at least one supercharge becomes a
space-time scalar. Twisting amounts to embedding the holonomy group
into the R-symmetry group.

Consider first the case when $\Sigma$ is flat. The holonomy group is
$U(1)_C$ in this case, so one needs to consider an embedding of
$U(1)_C$ into $SU(2)_R\times U(1)_N$. The two most obvious choices
are to identify $U(1)_C$ with the maximal torus of $SU(2)_R$ or with
$U(1)_N$. In the first case, the adjoint field $\phi$ remains a
0-form, while in the second case it becomes a 1-form on $C$ (with
values in the adjoint representation of $G$). Note that the first
twist is defined for an arbitrary (not necessarily superconformal)
$\cN=2$ theory, while the second one makes sense only in the
superconformal case. In what follows we will refer to the first and
second possibilities as the $\a$-twist and $\b$-twist, respectively.
Note also that a twist along $C$ makes sense whether the metric on
$\Sigma$ has Euclidean or Minkowski signature.

Let us begin with the $\a$-twisted theory. In the untwisted theory,
the bosonic fields are the gauge field $A_\mu$, the complex adjoint
scalar $\phi$ (which is the lowest component of the superfield
$\Phi$) and a pair of complex scalars $q$ and $\tq$ in
representations $R$ and $R^\vee$ of $G$ (these are the lowest
components of $Q$ and $\tQ$, respectively). The $U(1)_R$ subgroup of
$SU(2)_R$ which we will use for twisting can be chosen so that both
$q$ and $\tq$ have $U(1)_R$-charge $-1$ (and $\phi$ has charge $0$).
Then in the $\a$-model $q$ and $\tq$ become sections of
$K_C^{-1/2}$, where $K_C$ is the canonical line bundle of $C$ (or
rather, its pullback to $\Sigma\times C$).

As for fermionic fields, in the untwisted theory we have gauginos
$\lambda_1$ and $\lambda_2$ (in the adjoint of $G$) and quark fields
$\psi$ and $\chi$ (in representations $R^\vee$ and $R$). All these
fields are Weyl fermions; the $U(1)_R$ charges of $\lambda_1$ and
$\lambda_2$ are $+1$ and $-1$, respectively, while the $U(1)_R$
charges of $\psi$ and $\chi$ are zero. There are also right-handed
spinors (quarks and gauginos) with opposite R-charges. In Minkowski
signature they are related to the left-handed ones by complex
conjugation; in Euclidean signature they are independent fields and
have to be considered separately. We will distinguish the
right-handed ``partners'' of left-handed fields with a bar.

In the $\a$-model $\psi$ and $\chi$ remain left-handed spinors.
Using complex structures on $C$ and $\Sigma$, we can think of
left-handed spinors as sections of
$$
S_-=K_\Sigma^{-1/2}\ot K_C^{1/2}+K_\Sigma^{1/2}\ot K_C^{-1/2}.
$$
Similarly, the right-handed spinors $\bpsi,\bchi$ are sections of
$$
S_+=K_\Sigma^{-1/2}\ot K_C^{-1/2}+K_\Sigma^{1/2}\ot K_C^{1/2}.
$$
On the other hand, the gauginos $\lambda_1$ and $\lambda_2$
become sections of the vector bundles
\begin{equation}\label{glua1}
K_\Sigma^{-1/2}\ot K_C+K_\Sigma^{1/2}\ot \cO_C
\end{equation}
and
\begin{equation}\label{glua2}
K_\Sigma^{-1/2}\ot \cO_C+K_\Sigma^{1/2}\ot K_C^{-1},
\end{equation}
tensored with the gauge bundle in the adjoint representation.
Their right-handed partners become sections of
$$
K_\Sigma^{-1/2}\ot K_C^{-1}+K_\Sigma^{1/2}\ot \cO_C
$$
and
$$
K_\Sigma^{-1/2}\ot \cO_C+K_\Sigma^{1/2}\ot K_C,
$$
also tensored with the adjoint gauge bundle.

In flat space-time the theory has eight complex supercharges which
can be assembled into two left-handed spinors and two right-handed
spinors. In the twisted theory, only those which are scalars on $C$
survive. The transformation properties of the supercharges with
respect to R-symmetries are identical to those of the gauginos, so
from eqs. (\ref{glua1},\ref{glua2}) we conclude that the $\a$-model
has four complex supercharges, two of which transform as spinors of
one chirality on $\Sigma$, and the other two transform as spinors of
opposite chirality. If $\Sigma$ has Minkowski signature, then for
each chirality the two supercharges are related by complex
conjugation. If in the Kaluza-Klein spirit we regard the twisted
theory as a field theory living on $\Sigma$, then it has $(2,2)$
supersymmetry.

Now let us perform the same analysis for the $\b$-twisted theory.
The scalar $\phi$ has $U(1)_N$ charge $+2$, so after twist it
becomes a section of $K_C$. The scalars $q$ and $\tq$ have zero
$U(1)_N$ charge, so they are unaffected by the $\b$-twist. Both
$\psi$ and $\chi$ have $U(1)_N$ charge $-1$, so they become sections
of the vector bundle
\begin{equation}\label{psib}
K_\Sigma^{-1/2}\ot \cO_C+K_\Sigma^{1/2}\ot K_C^{-1}
\end{equation}
tensored with the gauge bundle in the representation $R^\vee$ or
$R$. Their right-handed partners become sections of
$$
K_\Sigma^{-1/2}\ot \cO_C+K_\Sigma^{1/2}\ot K_C
$$
tensored with the gauge bundle in the representation $R$ or
$R^\vee$. Both gauginos have $U(1)_N$ charge $+1$, so they become
sections of the vector bundle
$$
K_\Sigma^{-1/2}\ot K_C+K_\Sigma^{1/2}\ot \cO_C
$$
tensored with the adjoint gauge bundle. Their right-handed
partners are sections of
$$
K_\Sigma^{-1/2}\ot K_C^{-1}+K_\Sigma^{1/2}\ot \cO_C
$$
tensored with the adjoint gauge bundle.

The $\b$-model has four complex supercharges which are spinors of
the same chirality on $\Sigma$. If $\Sigma$ has Minkowski signature,
they are pairwise related by complex conjugation, so there are four
independent real supercharges. This means that the $\b$-model, when
regarded as a field theory on $\Sigma$, has chiral (left-handed)
$(4,0)$ supersymmetry. We will see later that under Kaluza-Klein
reduction it becomes a $(4,0)$ sigma-model on $\Sigma$ whose target
is the moduli space of Hitchin equations on $C$, and with
right-handed fermions taking values in a certain vector bundle over
the Hitchin moduli space.

\subsection{Partial twist along $\Sigma$}

Now let us consider the situation where both $\Sigma$ and $C$ are
curved. Then in order to preserve at least one fermionic symmetry
one has to perform a further twist of the holonomy group
$U(1)_\Sigma$. Another motivation to twist the theory along $\Sigma$
is to simplify the dependence of the theory on the metric: we will
see below that after $\Sigma$-twist the theory becomes independent
of the \Ka\ forms of both $C$ and $\Sigma$ (although it still
depends on their complex structures). Thus the theory twisted both
along $C$ and $\Sigma$ is equivalent to its Kaluza-Klein reduction
along $C$ or $\Sigma$.

For either $\a$ or $\b$ models, there are many possibilities for
twisting along $\Sigma$. In order to remain as close as possible to
the GL twist, we will twist the $\a$-model by identifying
$U(1)_\Sigma$ with $U(1)_N$. We will call this the $\a'$-model. It
is also natural to twist the $\b$-model by identifying $U(1)_\Sigma$
with the maximal torus of $SU(2)_R$. We will call this the
$\b'$-model. Obviously, $\a'$ and $\b'$-models are related by the
exchange of $C$ and $\Sigma$, so from now on we will only consider
the $\b'$-model. Let us also mention that if we use the maximal
torus of $SU(2)_R$ to twist both along $C$ and $\Sigma$, then we get
nothing but the Witten-Donaldson topological twist of the $\cN=2$
theory, in the special case when the four-manifold on which the
theory lives is taken to be $C\times \Sigma$.

Let us now describe the field content of the $\b'$-model. The
bosonic fields are the gauge field $A_\mu$, the adjoint Higgs field
$\phi$ which is a section of $K_C$, and the squark fields $q$ and
$\tq$ which after the $\Sigma$-twist become sections of
$K_\Sigma^{-1/2}$ (tensored with the gauge bundle in representation
$R^\vee$ or $R$). That is, the squark fields becomes chiral spinors
on $\Sigma$. The quark fields $\psi$ and $\chi$ are unaffected by
the $\Sigma$-twist and remain sections of the vector bundle
$$
K_\Sigma^{-1/2}\ot \cO_C+K_\Sigma^{1/2}\ot K_C^{-1}.
$$
That is, they remain spinors along $\Sigma$. Their right-handed
partners are sections of
$$
K_\Sigma^{-1/2}\ot \cO_C+K_\Sigma^{1/2}\ot K_C
$$
On the other hand, the gauginos $\lambda_1$ and $\lambda_2$ become
sections of the vector bundles
$$
\cO_\Sigma\ot K_C+K_\Sigma\ot \cO_C
$$
and
$$
K_\Sigma^{-1}\ot K_C+\cO_\Sigma\ot \cO_C.
$$
That is, they become differential forms along both $\Sigma$ and
$C$. Their right-handed partners become sections of
$$
K_\Sigma^{-1}\ot K_C^{-1}+\cO_\Sigma\ot\cO_C
$$
and
$$
\cO_\Sigma\ot K_C^{-1}+K_\Sigma\ot\cO_C.
$$
On a general $\Sigma$ there are two fermionic symmetries (one from
left-handed supercharge and one from right-handed supercharge in
four dimensions). These two supercharges have the same chirality
on $\Sigma$ and opposite chirality on $C$.

All the models described so far contain spinor fields on $C$ or
$\Sigma$ and therefore depend on the choice of spin structure. We
can avoid this by modifying the twist further. Namely, we add to the
generator of $U(1)_C$ or $U(1)_\Sigma$ a multiple of a generator of
$U(1)_B$. This has the effect of modifying the spins of the fields
in a way which depends on their $U(1)_B$ charge. The fields in the
$\cN=2$ vector multiplet are unaffected, but we can use this new
freedom to make the spins of all fields in the hypermultiplet to be
integral.

In the case of the $\b'$-model, we modify the generator of
$U(1)_\Sigma$ so as to make $\tq$ a 0-form on both $C$ and $\Sigma$.
We will refer to the resulting theory as the $\b''$-model. The field
content of $\a,\b,\b',$ and $\b''$ models is listed in Table
\ref{tableone}. In what follows, we will focus on the
$\beta''$-model, as its properties most closely resemble the
properties of the GL-twisted $\cN=4$ theory. Note that even if we
take the hypermultiplet to be in the adjoint representation of $G$,
the $\beta''$-model does not reduce to the GL-twisted theory.
Instead it reduces to one of the more general twisted theories
mentioned in section 5.1 of \cite{KW}.

\begin{table}\label{tableone}
\begin{tabular}{l|rr|rr|rr|rr}
  & \multicolumn{2}{c|}{$\a$} & \multicolumn{2}{c|}{$\b$} & \multicolumn{2}{c|}{$\b'$} &
 \multicolumn{2}{c}{$\b''$}\\ \hline
    Field   & $U(1)_C$ & $U(1)_\Sigma$ & $U(1)_C$ & $U(1)_\Sigma$ & $U(1)_C$ & $U(1)_\Sigma$ & $U(1)_C$ &
       $U(1)_\Sigma$\\ \hline
  $\phi$          & $0$  & $0$  & $2$  & $0$  & $2$  & $0$  & $2$  & $0$\\
  $\lambda_{1+}$  & $2$  & $-1$ & $2$  & $-1$ & $2$  & $0$  & $2$  & $0$ \\
  $\lambda_{1-}$  & $0$  & $1$  & $0$  & $1$  & $0$  & $2$  & $0$  & $2$ \\
  $\lambda_{2+}$  & $0$  & $-1$ & $2$  & $-1$ & $2$  & $-2$ & $2$  & $-2$ \\
  $\lambda_{2-}$  & $-2$ & $1$  & $0$  & $1$  & $0$  & $0$  & $0$  & $0$ \\
  $\blambda^1_+$ & $-2$ & $-1$ & $-2$ & $-1$ & $-2$ & $-2$ & $-2$ & $-2$ \\
  $\blambda^1_-$ & $0$  & $1$  & $0$  & $1$  & $0$  & $0$  & $0$  & $0$ \\
  $\blambda^2_+$ & $0$  & $-1$ & $-2$ & $-1$ & $-2$ & $0$  & $-2$ & $0$ \\
  $\blambda^2_-$ & $2$  & $1$  & $0$  & $1$  & $0$  & $2$  & $0$  & $2$ \\
  $q$             & $-1$ & $0$  & $0$  & $0$  & $0$  & $-1$ & $0$  & $-2$ \\
  $\tq$           & $-1$ & $0$  & $0$  & $0$  & $0$  & $-1$ & $0$  & $0$ \\
  $\psi_+$        & $1$  & $-1$ & $0$  & $-1$ & $0$  & $-1$ & $0$  & $-2$ \\
  $\psi_-$        & $-1$ & $1$  & $-2$ & $1$  & $-2$ & $1$  & $-2$ & $0$ \\
  $\chi_+$        & $1$  & $-1$ & $0$  & $-1$ & $0$  & $-1$ & $0$  & $0$ \\
  $\chi_-$        & $-1$ & $1$  & $-2$ & $1$  & $-2$ & $1$  & $-2$ & $2$ \\
  $\bpsi_+$       & $-1$ & $-1$ & $0$  & $-1$ & $0$  & $-1$ & $0$  & $0$ \\
  $\bpsi_-$       & $1$  & $1$  & $2$  & $1$  & $2$  & $1$  & $2$  & $2$ \\
  $\bchi_+$       & $-1$ & $-1$ & $0$  & $-1$ & $0$  & $-1$ & $0$  & $-2$ \\
  $\bchi_-$       & $1$  & $1$  & $2$  & $1$  & $2$  & $1$  & $2$  & $0$
\end{tabular}
\caption{The field content of various twisted models. The subscripts
$\pm$ on spinor fields refer to upper and lower components.}
\end{table}

\subsection{BRST transformations}\label{sec:BRST}

In the $\beta''$-model there are two independent BRST charges which
anticommute and square to zero. If we denote them by $Q_\ell$ and
$Q_r$, then the most general BRST charge is
$$
Q=u Q_\ell +v Q_r.
$$
Of course, the resulting theory depends only on the ratio $t=v/u$.
This is similar to the GL twist of $\cN=4$ super-Yang-Mills theory.
But there is also an important difference: in the case considered
here, there is a $U(1)$ symmetry with respect to which $Q_\ell$ and
$Q_r$ transform with opposite weights, so variation of the phase of
$t$ can be undone by a global symmetry transformation. In other
words, the phase of $t$ is irrelevant. The $U(1)$ symmetry in
question is $U(1)_N$. Further, the BRST charge depends
holomorphically on $t$, and we will see below that the same is true
about the action of the twisted theory. This implies that
$t$-dependence is trivial (can be undone by a symmetry
transformation) provided $t\neq 0,\infty$. We will see below that
properties of the $\beta''$-model for $t=0,\infty$ are very
different form those at other values of $t$.

As explained in \cite{KW}, under S-duality the left-handed and
right-handed supersymmetries are multiplied by opposite phases,
which is equivalent to saying that under S-duality the parameter $t$
is multiplied by a phase. But since the dependence on the phase of
$t$ is essentially trivial, we conclude that S-duality maps the
$\beta''$-model to itself (i.e. $t$ is unchanged). This is unlike
the GL-twist, where the phase of $t$ affects the theory in a
nontrivial way.

It is straightforward to work out BRST transformations of all the
fields. Local complex coordinates on $\Sigma$ and $C$ will be
denoted $w$ and $z$ respectively. Let us begin with the $\cN=2$
vector multiplet. The Higgs field is a section of $K_C$ and will be
denoted $\phi_z$. The components of $\lambda_1$ will be denoted
$\lambda_z, \lambda_w$, the components of $\lambda_2$ will be
$\lambda_{\bw z},\lambda_{z\bz}$, the components of $\bar\lambda_1$
will be $\bar\lambda_{\bw\bz},\bar\lambda_{z\bz}$, the components of
$\bar\lambda_2$ will be $\bar\lambda_\bz,\bar\lambda_w$. The BRST
variations of bosons are
\begin{align*}
\delta A_z &= -i\bar\xi\lambda_z, & \delta A_\bz &= -i\xi\bar\lambda_\bz\\
\delta A_w &=i\xi\bar\lambda_w+i\bar\xi\lambda_w , & \delta A_\bw &= 0\\
\delta \phi_z &=\xi\lambda_z, & \delta\phi_\bz
&=\bar\xi\bar\lambda_\bz
\end{align*}
The BRST variations of fermions are
\begin{align*}
\delta\lambda_z &=0,\\
\delta\lambda_w &=4\xi q_w T \tq^\dag,\\
\delta\lambda_{\bw z}&=-4\xi F_{\bw z}-4i\bxi D_\bw\phi_z,\\
\delta\lambda_{z\bz} &= 2\left(-i\xi[\phi_z,\phi_\bz]+\xi
F_{z\bz}-2i\bxi D_\bz\phi_z\right) -2\xi g^{w\bw}g_{z\bz}\left(F_{w\bw}+i \mu_{w\bw}\right),\\
\delta\blambda_\bz &=0,\\
\delta\blambda_w &=-4\bxi q_w T \tq^\dag,\\
\delta\blambda_{\bw\bz}&= -4\bxi F_{\bw\bz}-4i\xi D_\bw\phi_\bz,\\
\delta\blambda_{z\bz} &= 2\left(i\bxi[\phi_z,\phi_\bz]-\bxi
F_{z\bz}-2i\xi D_z\phi_\bz\right)-2\bxi g^{w\bw}g_{z\bz}
\left(F_{w\bw}+i\mu_{w\bw}\right).
\end{align*}
Here we denoted
$$
\mu_{w\bw}=q_w T q_\bw-2 g_{w\bw} \tq T \tq^\dag.
$$
and $T$ denotes the intertwiner between $R\otimes R^\vee$ and the
adjoint representation.

{}From these formulas, we see that the fields
$\lambda_z,\blambda_\bz,\lambda_w,\blambda_w$ have ghost number
$+1$, while the fields $\lambda_{z\bz},\blambda_{z\bz},\lambda_{\bw
z},\blambda_{\bw\bz}$ have ghost number $-1$. (By definition, BRST
transformation increases the ghost number by $1$).

Now let us turn to the matter fields. The BRST variations of
squarks are
\begin{align*}
\delta q_w&=0, & \delta q_\bw &= \xi \psi_\bw + \bxi \bchi_\bw,\\
\delta \tq & =\frac{i}{2}g^{w\bw}\left(\xi\chi_{w\bw} -
\bxi\bpsi_{w\bw}\right), & \delta \tq^\dag &=0.
\end{align*}

The BRST transformations of quark fields are
\begin{align*}
\delta\psi_\bw &= -4\bxi D_\bw \tq^\dag, & \delta\psi_\bz
&=-4\left(\bxi
D_\bz-\xi \phi_\bz\right) \tq^\dag,\\
\delta\chi_{w\bw} &= -4i\bxi D_\bw q_w, & \delta\chi_{w\bz}
&=-4i\left(\bxi D_\bz q_w + \xi q_w \phi_\bz\right),\\
\delta \bpsi_{w\bw} &= -4i\xi D_\bw q_w    & \delta\bpsi_{wz} &= -4i
\left(\xi D_z q_w+\bxi q_w\phi_z\right),  \\
\delta \bchi_\bw & = 4\xi D_\bw \tq^\dag & \delta\bchi_z &= 4
\left(\xi D_z -\bxi \phi_z\right) \tq^\dag .
\end{align*}

The ghost number assignments of matter fields are not unique,
because of the freedom to add to the ghost number a multiple of
$U(1)_B$ charge. One fairly natural choice is to require $q_\bw$ and
$\tq$ to have the same ghost number $gh$. Then the ghost numbers of
all matter fields are uniquely determined: $q_w$ and $\tq^\dag$ have
$gh=1$, $q_\bw$ and $\tq$ have $gh=-1$, and all fermionic matter
fields have $gh=0$. As we will see below, this definition of the
ghost number is natural from the viewpoint of the effective field
theory on $\Sigma$. {}From the viewpoint of the effective field theory
on $C$, it is more natural to let $q_w$ and $q_\bw$ to have $gh=0$.
Then $\tq$ and $\tq^\dag$ have ghost numbers $-2$ and $2$,
respectively. As for fermionic matter fields, all 1-form fermions
have $gh=1$, while all 2-form fermions have $gh=-1$.

The above formulas describe how fields transform under both
$Q_\ell$ and $Q_r$. The transformation law under the BRST charge
$Q$ is obtained by taking $\bxi$ to be proportional to $\xi$:
$$
\bxi=-it\xi,\quad t\in\CC\cup \{\infty\}
$$
We will denote the corresponding BRST variation $\delta_t$.

The BRST transformation $\delta_t$ is almost off-shell nilpotent:
without using equations of motion, the equation $\delta_t^2\Phi=0$
holds for all fields except $\lambda_{z\bz}$ and $\blambda_{z\bz}$,
but if we use the equations of motion for these two fields we get
$\delta_t^2\Phi=0$ for all fields. In fact, one can do slightly
better by introducing the linear combinations
$$
\Upsilon_{z\bz}=\lambda_{z\bz}-it^{-1}\blambda_{z\bz},\quad
\Omega_{z\bz}=\lambda_{z\bz}+it^{-1}\blambda_{z\bz},
$$
because then we have $\delta_t^2\Upsilon_{z\bz}=0$, and therefore
$\delta_t$ fails to be off-shell nilpotent only on $\Omega_{z\bz}$.

To check the nilpotency of $\delta_t$, it is very useful to note
that the complex connection $\cA$ defined by
$$
\cA_z=A_z+t\phi_z,\quad \cA_\bz=A_\bz-t^{-1}\phi_\bz,\quad
\cA_w=A_w,\quad \cA_\bw=A_\bw
$$
is BRST-invariant for all values of $t$. If we denote by $\cF$ the
curvature 2-form of this connection, then the BRST-variations of
gauginos take the form
\begin{align}\label{fermisimple}
\delta_t\lambda_{\bw z} &=-4\cF_{\bw z}, \\
\delta_t\blambda_{\bw\bz}&=4it\cF_{\bw\bz},\\
\delta_t\Upsilon_{z\bz} &=4\cF_{z\bz},\\
\delta_t\Omega_{z\bz}
&=4\left(t^{-1}D_z\phi_\bz-tD_\bz\phi_z\right)-4g^{w\bw}g_{z\bz}\left(F_{w\bw}+i\mu_{w\bw}\right).
\end{align}
The off-shell nilpotency of $\delta_t$ when acting on $\lambda_{\bw
z},\lambda_{\bw\bz},$ and $\Upsilon_{z\bz}$ is manifest.

In order to write down BRST-invariant actions, it is very convenient
to make the BRST transformation nilpotent off-shell by introducing
suitable auxiliary fields. In the present case we need just one such
field which we call $H_{z\bz}$. We will change the action of
$\delta_t$ on $\Omega_{z\bz}$ and postulate
$$
\delta_t\Omega_{z\bz}=H_{z\bz}, \quad \delta_t H_{z\bz}=0.
$$
If we want the new BRST transformations to be equivalent on-shell to
the old ones, we have to ensure that on-shell $H_{z\bz}$ is equal to
the old BRST variation of $\Omega_{z\bz}$, i.e. we want the equation
of motion for $H_{z\bz}$ to read
$$
H_{z\bz}=4\left(t^{-1}D_z\phi_\bz-tD_\bz\phi_z\right)-4g^{w\bw}g_{z\bz}\left(F_{w\bw}+i\mu_{w\bw}\right).
$$

The modified BRST transformations of the fields in the $\cN=2$
vector multiplet enjoy the nice property that they are independent
of the K\"ahler forms $g_{w\bw}$ and $g_{z\bz}$. This is almost true
for the BRST transformations of the hypermultiplet fields: the only
offender is $\tq$. We can easily repair this by working with the
following linear combinations of the quark fields $\chi_{w\bw}$ and
$\bpsi_{w\bw}$:
\begin{equation}\label{eta}
\varrho=\frac{1}{2}g^{w\bw}\left(\chi_{w\bw}+it\bpsi_{w\bw}\right),\quad
\eta_{w\bw}=\chi_{w\bw}-it\bpsi_{w\bw}.
\end{equation}
Then we have
\begin{align*}
\delta_t\tq&=i\varrho,\\
\delta_t\varrho&=0,\\
\delta_t\eta_{w\bw}&=-8t D_\bw q_w.
\end{align*}

\subsection{Action and properties of the twisted theory}\label{action}

Recall that the action of a topologically twisted theory typically
can be written as a ``topological'' term plus a BRST-exact term. The
``topological'' term is independent of the metric and is a
locally-constant function in the space of fields. This implies that
correlators of BRST-closed operators are independent of the metric.

The $\b''$-model turns out to behave somewhat differently. We will
see that for $t\neq 0,\infty$ its action can be written as a
BRST-exact piece plus a fermionic piece which is BRST-closed but not
BRST exact. The fermionic piece is not ``topological'' in the sense
that it is not a total derivative, but it is still independent of
the \Ka\ forms of $C$ and $\Sigma$. This implies that correlators of
BRST-closed operators are independent of the \Ka\ structure,
although they may still depend on the complex structure. We will
also see that the $\b''$-model is independent of the gauge coupling
$e^2$ and the theta-angle.

For $t=0,\infty$  in addition to the fermionic piece there is also a
more traditional topological term proportional to the instanton
number, so the properties of the theory are rather different. In
particular for $t=0$ (resp. $t=\infty$) the observables of the
theory may have holomorphic (resp. anti-holomorphic) dependence on
the complexified gauge coupling $\tau$.

The action of the twisted $\cN=2$ Yang-Mills theory on $\CC^2$ with
a flat product metric $ds^2=g_{z\bz}dz\otimes d\bz+g_{w\bw}dw\otimes
d\bw$ is
\begin{align*}
I_{YM}&=\frac{1}{e^2}\int d^2z\, d^2w\,{\sqrt g}\, \left[{\rm
Tr}\left(\frac{1}{2}F_{\mu\nu}F^{\mu\nu}+2g^{z\bz}D_\mu\phi_zD^\mu\phi_\bz+
\left(g^{z\bz}\right)^2[\phi_z,\phi_\bz]^2
+\left(g^{w\bw}\mu_{w\bw}\right)^2 \right.\right.\\
&\left.\left. +8g^{w\bw} (\tq T q_\bw)(q_w T\tq^\dag)\right)
+4D_\mu\tq D^\mu \tq^\dag+2g^{w\bw}D_\mu q_w D^\mu q_\bw
+4g^{z\bz}\tq\{\phi_z,\phi_\bz\}\tq^\dag\right.\\
&\left. +2g^{z\bz}g^{w\bw}q_w
 \{\phi_z,\phi_\bz\} q_\bw
 -\Tr
 \left(ig^{w\bw}g^{z\bz}\blambda_{\bw\bz}\left(D_w\lambda_z+D_z\lambda_w\right)
+ig^{z\bz}\blambda_{z\bz}\left(g^{z\bz}D_\bz\lambda_z-g^{w\bw}D_\bw\lambda_w\right)\right.\right.\\
 &\left.\left. +ig^{z\bz}\blambda_\bz\left(g^{w\bw}D_w\lambda_{\bw z}+g^{z\bz} D_z\lambda_{z\bz}\right)
 +ig^{w\bw}g^{z\bz}\blambda_w\left(D_\bz\lambda_{\bw
 z}-D_\bw\lambda_{z\bz}\right)\right.\right.\\
&\left.\left.
+ig^{z\bz}\phi_\bz\left(g^{z\bz}\{\lambda_z,\lambda_{z\bz}\}-g^{w\bw}\{\lambda_w,\lambda_{\bw
z}\}\right)
 +ig^{z\bz}\phi_z\left(g^{z\bz}\{\blambda_\bz,\blambda_{z\bz}\}-g^{w\bw}\{\blambda_w,\blambda_{\bw\bz}\}
 \right)\right)\right.\\
 & \left. -ig^{w\bw}\bpsi_{w\bw}\left(g^{w\bw} D_w\psi_\bw+g^{z\bz} D_z\psi_\bz\right)
 -ig^{w\bw}g^{z\bz}\bpsi_{wz}\left(D_\bz\psi_\bw-D_\bw\psi_\bz\right)\right.\\
 &\left. -ig^{w\bw}\chi_{w\bw}\left(g^{z\bz}D_\bz\bchi_z+g^{w\bw}D_w\bchi_\bw\right)
 +ig^{w\bw}g^{z\bz}\chi_{w\bz}\left(D_\bw\bchi_z-D_z\bchi_\bw\right)\right.\\
 & \left. -ig^{w\bw}g^{z\bz}q_w\left(\lambda_{z\bz}\psi_\bw-\lambda_{\bw z}\psi_\bz\right)
 -ig^{w\bw}\left(g^{z\bz}\bpsi_{wz}\blambda_\bz-g^{w\bw}\bpsi_{w\bw}\blambda_w\right)q_\bw\right.\\
 &\left. -g^{z\bz}g^{w\bw}\left(\chi_{w\bz}\lambda_{\bw z}-\chi_{w\bw}\lambda_{z\bz}\right)\tq^\dag
 +2\tq \left(g^{w\bw}\blambda_w\bchi_\bw-g^{z\bz}\blambda_\bz\bchi_z\right)\right.\\
 &\left. -ig^{w\bw}\left(g^{z\bz}\chi_{w\bz}\lambda_z-g^{w\bw}\chi_{w\bw}\lambda_w\right)q_\bw
 -ig^{z\bz}g^{w\bw}q_w\left(\blambda_{z\bz}\bchi_\bw-\blambda_{\bw\bz}\bchi_z\right)\right.\\
 &\left. -2\tq\left(g^{w\bw}\lambda_w\psi_\bw-g^{z\bz}\lambda_z\psi_\bz\right)
 +g^{z\bz}g^{w\bw}\left(\bpsi_{wz}\blambda_{\bw\bz}-\bpsi_{w\bw}\blambda_{z\bz}\right)\tq^\dag\right.\\
 &\left. -ig^{z\bz}g^{w\bw}\left(\chi_{w\bz}\phi_z\psi_\bw-\chi_{w\bw}\phi_z\psi_\bz
 +\bpsi_{wz}\phi_z\bchi_\bw-\bpsi_{w\bw}\phi_\bz\bchi_z\right)
 \right]
\end{align*}

We let
\begin{align*}
\Psi_1 &=-\frac{1}{2} g^{z\bz}g^{w\bw}\Tr\,\lambda_{\bw z}
\left(F_{w\bz}+t^{-1} D_w\phi_\bz\right),\\
\Psi_2&=-\frac{it^{-1}}{2}g^{z\bz}g^{w\bw}\Tr\,\blambda_{\bw\bz}\left(F_{wz}-t
D_w\phi_z\right),\\
\Psi_3&=-\frac{1}{4}\left(g^{z\bz}\right)^2\Tr\,\Upsilon_{z\bz}\left(F_{z\bz}-i[\phi_z,\phi_\bz]+t^{-1}D_z
\phi_\bz + t
D_\bz\phi_z\right),\\
\Psi_4&=\frac{1}{16}\left(g^{z\bz}\right)^2\Tr\,
\Omega_{z\bz}\left(H_{z\bz}-8\left(t^{-1}D_z\phi_\bz-tD_\bz\phi_z\right)+
8g^{w\bw}g_{z\bz}\left(F_{w\bw}+i\mu_{w\bw}\right)\right).\\
\Psi_5&=g^{w\bw}D_w\tq\left(\bchi_\bw-it^{-1}\psi_\bw\right)-it^{-1}g^{z\bz}\left(D_z\tq+it\tq
\phi_z\right)\psi_\bz+g^{z\bz}\left(D_\bz\tq-it^{-1}\tq\phi_\bz\right)\bchi_z,\\
\Psi_6&=-\frac{t^{-1}}{2}\left(g^{w\bw}\right)^2 \eta_{w\bw} D_w
q_\bw-\frac{t^{-1}}{2}g^{w\bw}g^{z\bz}\chi_{w\bz}\left(D_z
q_\bw-it\phi_z q_\bw\right)+\frac{i}{2}g^{w\bw}
g^{z\bz}\bpsi_{wz}\left(D_\bz q_\bw+it^{-1} \phi_\bz q_\bw\right),\\
\Psi_7&=g^{w\bw}\tq \left(\lambda_w-it^{-1}\blambda_w\right)q_\bw.
\end{align*}

Consider the following ansatz for a BRST-invariant action:
$$
I_0=\frac{1}{e^2}\sum_{i=1}^7 \int d^2z\, d^2w\, {\sqrt g}\,
\delta_t \Psi_i
$$
Its bosonic part is $t$-independent and in flat space-time is equal
to the bosonic part of $I_{YM}$. But its fermionic part is
$t$-dependent and does not coincide with the fermionic part of
$I_{YM}$. To write down the difference between the fermionic parts
of $I_{YM}$ and $I_0$ in a compact form, we will make use of the
remaining letters of the Greek alphabet, except the really strange
ones. We define
\begin{align}\label{forms}
\Upsilon &=\Upsilon_{z\bz} dz\wedge d\bz \\
\sigma & =\bchi_z dz - i t^{-1} \psi_\bz d\bz,\nn\\
\Psi_w &=-it \bpsi_{wz} dz + \chi_{w\bz} d\bz,\nn\\
\zeta_\bw &=\lambda_{\bw z} dz+ i t^{-1}\blambda_{\bw\bz} d\bz,\nn\\
\beta_\bw &=\bchi_\bw-i t^{-1}\psi_\bw,\nn\\
\Lambda_w &= i\blambda_w-t\lambda_w\nn
\end{align}
Thus $\Upsilon$ is a 2-form on $C$, $\sigma,\Psi_w,$ and $\zeta_\bw$
are 1-forms on $C$, and $\beta_\bw,\Lambda_w$ are 0-forms on $C$.
Let $\cD_C$ denote the covariant differential along $C$; it maps
$k$-forms on $C$ to $k+1$-forms. Then the difference between the
fermionic parts of $I_{YM}$ and $I_0$ is
\begin{align*}
I_1&=\frac{1}{2e^2}\int_{C\times\Sigma} dw\wedge d\bw\
\left(\Tr\left(\zeta_{\bw} \cD_C \Lambda_w-\Upsilon
D_\bw\Lambda_w\right)+2i\Psi_w
D_\bw \sigma - i\Psi_w \cD_C\beta_\bw\right.\\
& \left. - i\eta_{w\bw} \cD_C \sigma - \eta_{w\bw} \Upsilon \tq^\dag
-t q_w \Upsilon \beta_\bw + 2t q_w\zeta_\bw \sigma-2\Psi_w\zeta_\bw
\tq^\dag \right)
\end{align*}
One can check that $I_1$ is BRST-invariant, as it should. Thus if we
want the twisted theory to be BRST-invariant and agree with the
$\cN=4$ SYM in flat space-time, we can take its action to be
\begin{equation}\label{actionfull}
I=I_0+I_1+I_\theta,
\end{equation}
where
$$
I_\theta=-\frac{i\theta}{8\pi^2}\int \Tr\ F\wedge F.
$$

We immediately observe that the action depends on the  \Ka\ forms of
$C$ and $\Sigma$ only through BRST-exact terms. Therefore the
correlators in the $\b''$-model cannot depend on the \Ka\ forms.

In fact, a stronger result holds: the theory is independent of the
complex structure of $C$, which makes it topological along $C$ and
holomorphic along $\Sigma$. Indeed, apart from the BRST-exact term
$I_0$, the action does not make reference to the decomposition of
the forms along $C$ into holomorphic and antiholomorphic components.
This is the reason we introduced linear combinations (\ref{forms})
in the first place.

Another important consequence of (\ref{actionfull}) is that after
rescaling all the fermions by a factor $e$ the action depends on $e$
only through BRST-exact terms. Thus correlators of BRST-invariant
operators are independent of $e$. This means that weak-coupling
computations in the $\b''$-model give exact results.

It turns out that for $t\neq 0,\infty$ varying the theta-angle also
does not affect the correlators. To see this, recall that the
path-integral of the twisted theory localizes on BRST-invariant
bosonic field configurations. We will refer to them as BPS
configurations. They are solutions of the equations
$\delta_t\Phi=0$, where $\Phi$ is any fermionic field. {}From
(\ref{fermisimple}) we see that BPS configurations satisfy
$$
\cF_{\bw z}=\cF_{\bw\bz}=\cF_{z\bz}=0,
$$
which implies $\cF\wedge\cF=0$. The connections $A$ and $\cA$ have
the same instanton number, hence BPS configurations are
automatically topologically trivial, and the correlators in the
$\b''$-model are independent of $\theta$.\footnote{If we allow
$C\times\Sigma$ to have boundaries, the difference between $\int
\cF^2$ and $\int F^2$ need not vanish, and the question of
$\theta$-dependence has to be re-examined.}

The preceding analysis does not apply to two exceptional cases $t=0$
and $t=\infty$. We will consider explicitly only the case
$t=\infty$, as $t=0$ is very similar. For $t=\infty$ (which
corresponds to setting $\xi=0$) the BRST transformations close
off-shell for all fields except $\blambda_{z\bz}$. So we introduce
an auxiliary field $G_{z\bz}$ and redefine the BRST transformation
on $\blambda_{z\bz}$ as
$$
\delta_\infty \blambda_{z\bz}=G_{z\bz}.
$$
We also let
$$
\delta_\infty G_{z\bz}=0
$$
and construct the action so that the equation of motion for
$G_{z\bz}$ be
$$
G_{z\bz}=2\left(i[\phi_z,\phi_\bz]-F_{z\bz}\right)-2g_{z\bz}g^{w\bw}\left(F_{w\bw}+i\mu_{w\bw}\right).
$$
If we also redefine
$$
\varrho=-\frac{i}{2}g^{w\bw}\bpsi_{w\bw}
$$
we note that the modified BRST transformations are independent of
both \Ka\ forms.

To write down the action at $t=\infty$ we define
\begin{align*}
\Xi_1 &= -g^{z\bz}g^{w\bw}\Tr\, \blambda_{\bw\bz} F_{wz} + i
g^{z\bz}g^{w\bw}\Tr\,
\lambda_{\bw z} D_w\phi_\bz,\\
\Xi_2
&=i(g^{z\bz})^2\lambda_{z\bz}D_z\phi_\bz+\frac{1}{4}\left(g^{z\bz}\right)^2\Tr\,
\blambda_{z\bz}\left(G_{z\bz}-4\left(i[\phi_z,\phi_\bz]-F_{z\bz}\right)+
4g_{z\bz}g^{w\bw}\left(F_{w\bw}+i\mu_{w\bw}\right)\right),\\
\Xi_3 &= -2g^{z\bz} D_z \tq\, \psi_\bz- 2 g^{w\bw}  D_w \tq\, \psi_\bw + ig^{z\bz}g^{w\bw}\chi_{w\bz}D_z q_\bw+i\left(g^{w\bw}\right)^2\chi_{w\bw} D_w q_\bw\\
\Xi_4 &= -2g^{z\bz} \tq\phi_\bz \bchi_z+i g^{z\bz}
g^{w\bw}\bpsi_{wz}\phi_\bz q_\bw -2g^{w\bw} \tq \blambda_w q_\bw .
\end{align*}
Then we consider the following BRST-exact action:
$$
I^\infty_0=\frac{1}{e^2}\int d^2z\, d^2 w \sqrt g \sum_{i=1}^4
\delta_\infty\Xi_i
$$
Its bosonic part is equal to the bosonic part of $I_{YM}$ plus a
topological term
$$
-\frac{1}{e^2}\int \Tr \, F\wedge F
$$
The fermionic part of $I^\infty_0$ is different from that of
$I_{YM}$, but if we define
\begin{align*}
I^\infty_1 &=\frac{i}{e^2}\int_{C\times\Sigma} dz\wedge d\bz\wedge
dw\wedge d\bw \left[\Tr\,\left(\lambda_{\bw z}
D_\bz\blambda_w-\lambda_{z\bz} D_\bw
\blambda_w-\phi_z\{\blambda_w,\blambda_{\bw\bz}\}\right)\right.\\
 &\left.+\bpsi_{wz}(D_\bz\psi_\bw-D_\bw\psi_\bz)+\left(\chi_{w\bw} D_\bz
 -\chi_{w\bz}D_\bw\right)\bchi_z - q_w\blambda_{\bw\bz}\bchi_z+i\bpsi_{wz}\blambda_{\bw\bz}\tq^\dag\right.\\
 &\left. +q_w\lambda_{z\bz}\psi_\bw-q_w\lambda_{\bw z}\psi_\bz+i\chi_{w\bw}\lambda_{z\bz}\tq^\dag
 -\chi_{w\bw}\phi_z\psi_\bz-i\chi_{w\bz}\lambda_{\bw z}\tq^\dag+\chi_{w\bz}\phi_z\psi_\bw\right]
\end{align*}
then in flat space-time we have an identity
\begin{equation}\label{actioninfty}
I_{YM}+I_{\theta}=I^\infty_0+I^\infty_1-\frac{i\btau}{4\pi}\int
\Tr\, F\wedge F
\end{equation}
Therefore we define the action of the $\b''$-model on a general
curved manifold of the form $C\times\Sigma$ to be the right-hand
side of (\ref{actioninfty}). By construction, $I^\infty_1$ is
BRST-invariant; it is also easy to check this explicitly.

Since $I^\infty_0$ is BRST-exact, and $I^\infty_1$ does not depend
on the \Ka\ forms of $C$ and $\Sigma$, we see that for $t=\infty$
the correlators in the twisted theory depend only on the complex
structures of $C$ and $\Sigma$. In fact, as we will see below,
correlators of gauge-invariant observables are holomorphic functions
of $w$ and $z$. (More precisely, they are holomorphic sections of
various holomorphic line bundles on $C\times\Sigma$). One can
summarize the situation by saying that for $t=\infty$ the
$\b''$-model is a holomorphic field theory on $C\times\Sigma$.

Note that after rescaling all the fermions by a factor $e$, the
action depends on $e^2$ and $\theta$ only through a combination
$$
-\frac{i\btau}{4\pi}\int \Tr\, F\wedge F.
$$
This implies that correlators are independent of $e^2$ and $\theta$
in perturbation theory, while on the nonperturbative level the
correlators are antiholomorphic functions of $\tau$.

For $t=0$, the situation is similar, except that the correlators
turn out to be holomorphic functions of $w$ and $\tau$ and
antiholomorphic functions of $z$.

\section{Kaluza-Klein reduction}\label{KK}

Our next goal is to interpret some features of the $\beta''$-model
in two-dimensional terms. To this end, we consider the limit in
which either $C$ or $\Sigma$ has vanishingly small size. In this
limit, one expects the twisted gauge theory to reduce to a field
theory on $\Sigma$ or $C$, respectively. In the case when $C$
shrinks to zero size, we will show that the effective 2d field
theory on $\Sigma$ is a half-twisted model whose target is the
Hitchin moduli space $\MH(G,C)$, and with right-moving fermions
taking values in a certain vector bundle over $\MH(G,C)$ which
depends on the matter content. In the case when $\Sigma$ shrinks to
zero size, we will show that the effective 2d field theory on $C$ is
a $B$-model whose target is the moduli space of the nonabelian
vortex equations.

\subsection{Reduction along $C$}

Consider first the limit in which the volume of $C$ goes to zero.
That is, we rescale
$$
g^{z\bz}\ra \eps^{-1}g^{z\bz}
$$
and take the limit $\eps\ra 0$. In this limit, the effective field
theory on $\Sigma$ must be a sigma-model whose target is the moduli
space of time-independent zero-energy configurations. We can replace
the zero-energy condition by the requirement of BRST-invariance. In
the limit $\eps\ra 0$ such configurations must be solutions of
\begin{align}
F_{z\bz}-i[\phi_z,\phi_\bz]-t D_\bz\phi_z&=0,\\
F_{z\bz}-i[\phi_z,\phi_\bz]-t^{-1}D_z\phi_\bz&=0,\\
\left(D_z+it\phi_z\right)\tq^\dag&=0,\\
\left(D_\bz-it^{-1}\phi_\bz\right)\tq^\dag&=0,\\
D_z q_w-itq_w\phi_z&=0,\\
D_\bz q_w+it^{-1} q_w \phi_\bz&=0.
\end{align}
The first two equations are equivalent to the Hitchin equations:
$$
F_{z\bz}-i[\phi_z,\phi_\bz]=0,\quad D_\bz\phi_z=0.
$$
The remaining equations require $\tq^\dag$ and $q_w$ to be
covariantly constant with respect to the connection 1-form
$$
\cA=\left(A_z+t\phi_z\right)dz+\left(A_\bz-t^{-1}\phi_\bz\right)d\bz
$$
Hitchin equations imply that this complex connection is flat.
Generically, it is also irreducible, and therefore the only
solution of the last four equations is $q_w=\tq^\dag=0$. That is,
all squark fields vanish. One also has to identify field
configurations which are related by a gauge transformation. The
net result is that the target space of the effective field theory
on $\Sigma$ is the moduli space of Hitchin equations $\MH(G,C)$.

This moduli space is well-known to be hyperk\"ahler, which means
that a left-right symmetric sigma-model with this target has $(4,4)$
supersymmetry. One can get a topological field theory on $\Sigma$ by
twisting this sigma-model. However, the effective field theory on
$\Sigma$ we get from the $\beta''$-model is not a twisted $(4,4)$
sigma-model. Rather, it is a twisted version of a $(4,0)$
sigma-model with the same target. To see this, recall that the
$\beta''$-model is a twisted version of the $\beta$-model. The
latter is a supersymmetric sigma-model with two complex supercharges
both of which have the same chirality. That is, it is a $(4,0)$
sigma-model. The BRST operator of the $\beta''$-model is a linear
combination of these two complex supercharges.

To completely specify a twisted $(4,0)$ theory, one has to specify a
holomorphic vector bundle on $\MH(G,C)$ (because the right-moving
fermions take values in this bundle) and a complex structure on
$\MH(G,C)$ (this tells us which linear combination of the two
supercharges one takes as the BRST charge).

To determine the complex structure, it is sufficient to examine
the equations satisfied by the BRST-invariant configurations.
These equations say that the following combinations must be
(covariantly) holomorphic functions of the variable $w$:
$$
A_z+t\phi_z,\quad A_\bz-t^{-1}\phi_\bz.
$$
Therefore the relevant complex structure on $\MH(G,C)$ is the one
for which these combinations are holomorphic coordinates. We will
call it $J_t$. By varying $t$, one can get an arbitrary complex
structure on $\MH(G,C)$.

The right-moving fermions on $\Sigma$ come from the quark fields
$\psi_\bz,\bchi_z$ and $\bpsi_{wz},\chi_{w\bz}$.\footnote{Thus it is
natural to define the ghost number so that all these fields have
$gh=0$.} The former are 0-forms on $\Sigma$, while the latter are
1-forms on $\Sigma$. It is sufficient to consider 0-form fermions,
since 1-form fermions automatically take values in the dual vector
bundle. The equations of motion for 0-form fermions in the limit
$\eps\ra 0$ read
$$
D_\bz\bchi_z-\phi_z\psi_\bz=0,\quad D_z\psi_\bz-\phi_\bz\bchi_z=0.
$$
Solutions of these equations define, roughly speaking, a vector
bundle $\cR$ over $\MH(G,C)$. (More precisely, it is a twisted
vector bundle, see below). To understand its geometric meaning and
to verify that it is holomorphic in complex structure $J_t$, recall
that we have previously defined a fermionic field $\sigma$ with
values in $R^\vee(E)\otimes \Omega^1_C$:
$$
\sigma=\bchi_z dz-it^{-1}\psi_\bz d\bz.
$$
The equations of motion for $\bchi_z$ and $\psi_z$ are equivalent to
\begin{equation}\label{derham}
\cD\sigma=0,\quad \cD\star_C\sigma=0.
\end{equation}

These equations clearly depend holomorphically on the complex flat
connection $\cA$ and define, roughly speaking, a holomorphic vector
bundle on the moduli space of flat $G_\CC$ connections on $C$. (More
precisely, it is a twisted vector bundle, see below). The latter
moduli space is isomorphic as a complex manifold to $\MH(G,C)$ in
complex structure $J_t$ (for any $t\neq 0,\infty$).

Let $\cR$ denote the holomorphic vector ``bundle'' defined by the
equations (\ref{derham}). The equations involve the Hodge operator
$\star_C$ on 1-forms, so it is not obvious that $\cR$ is independent
of the complex structure on $C$. Nevertheless, this must be so,
since we have shown that the $\b''$-model is independent of it. To
see how this comes about (at least, outside of a set of a high
codimension on $\MH(G,C)$), let us consider dropping the second
equation in (\ref{derham}) and identifying solutions whose
difference is $\cD_C$-exact. This gives a holomorphic vector
``bundle'' on $\MH(G,C)$ which we call $\cR'$. $\cR'$ is manifestly
independent of the complex structure on $C$. If in each equivalence
class there is a unique representative satisfying the equation that
we dropped, then $\cR'$ is isomorphic to $\cR$, and this would prove
that $\cR$ is also independent of the complex structure.

As usual, it is much easier to demonstrate uniqueness than
existence, and we will only argue the former. Suppose $\sigma$
satisfies both equations in (\ref{derham}) and has the form
$\sigma=\cD_C\varrho$ for some section $\varrho$ of the bundle
$R^\vee(E)$. Then $\varrho$ satisfies the second order equations
$$
D_\bz D_z\varrho-\phi_z\phi_\bz\varrho=0,\quad D_z
D_\bz\varrho-\phi_\bz\phi_z\varrho=0.
$$
We would like to show that $\varrho$ vanishes identically. To this
end we multiply, say, the second equation by $\varrho^\dag$ and
integrate over $C$. The resulting identity implies
$D_\bz\varrho=0,\quad \phi_z\varrho=0$. If the pair $(E,\varphi)$ is
stable, then these equations have only the trivial solution. This
proves the desired result (uniqueness) at least outside the strictly
semistable locus on $\MH(G,C)$.

Let us now explain why $\cR'$ is a twisted vector bundle. Since for
$t\neq 0,\infty$ $\MH(G,C)$ in complex structure $J_t$ is isomorphic
to the moduli space of flat $G_\CC$ connections on $C$, the product
$\MH(G,C)\times C$ carries a universal twisted flat principal
$G_\CC$-bundle, which we denote $\cE$. As explained in \cite{KW},
$\cE$ is twisted by a pull-back of a certain class $\zeta\in
H^2(\MH,Z(G))$. Given an irreducible representation $R$ of $G$, we
can associate to $\cE$ a twisted flat vector bundle $R^\vee(\cE)$ on
$\MH(G,C)\times C$; it is twisted by a class $R^\vee(\zeta)\in
H^2(\MH,U(1))$. If $R$ is reducible, then $R^\vee(\cE)$ is a sum of
twisted flat vector bundles. The vector bundle $\cR\simeq \cR'$ is
the 1-st fiberwise de Rham cohomology of $R^\vee(\cE)$ with respect
to the projection to $\MH(G,C)$.

At first sight, the appearance of twisted bundles seems worrisome,
because the path integral over fermions only makes sense if the
fermions take values in an ordinary bundle over the worldsheet
$\Sigma$. The potential paradox is resolved by recalling that for a
simply-connected gauge group such as $SU(N)$ the path-integral on
$S^1\times {\tilde S}^1\times C$ involves summation over all
electric fluxes on ${\tilde S}^1$ (and $C$). As explained in
\cite{KW}, section 7, from the point of view of the effective
sigma-model on $\Sigma$ summing over electric fluxes along ${\tilde
S}^1$ is the same as including only those maps from
$\Sigma=S^1\times {\tilde S}^1$ to $\MH(G,C)$ for which the
pull-back of $\zeta$ is trivial. The pull-back of $\cR$ by such a
map is an ordinary bundle on $\Sigma$.

Note that the vector bundle $\cR$ has a holomorphic structure with
respect to complex structure $J_t$ for arbitrary $t$. A vector
bundle on a hyperk\"ahler manifold can be holomorphic in all complex
structures simultaneously only if its first Chern class vanishes.
Hence we have $c_1(\cR)=0$, and the right-handed fermion number is
conserved. Of course, one can also show that $c_1(\cR)=0$ more
directly by using the index theorem for families.

\subsection{Reduction along $\Sigma$}

Now consider the limit in which the volume of $\Sigma$ goes to zero.
More precisely, we rescale
$$
g^{w\bw}\ra \eps^{-1}g^{w\bw}
$$
and take the limit $\eps\ra 0$. The resulting effective field theory
on $C$ turns out to be rather different from that on $\Sigma$: it is
a topological sigma-model (B-model) whose target is a certain
K\"ahler manifold $\MV(G,R,\Sigma)$. To see this, let us temporarily
undo the twist along $C$, or equivalently consider the
$\alpha$-twisted theory with $\Sigma$ and $C$ exchanged. This theory
has two complex supercharges of opposite chirality on $C$. That is,
it has $(2,2)$ supersymmetry. The BRST charge of the $\beta''$-model
is a linear combination of these two supercharges. If $t\neq
0,\infty$, then the $\beta''$-model reduces to a topologically
twisted $(2,2)$ field theory on $C$; otherwise the BRST charge is
purely left-moving or right-moving on $C$, and the $\beta''$-model
reduces to a half-twisted $(2,2)$ field theory on $C$.

To determine the target space of this $(2,2)$ sigma-model, we
consider BRST-invariant configurations in the limit $\eps\ra 0$. The
resulting equations for bosonic fields read
\begin{align*}
F_{w\bw}+i q_w T q_\bw&=0,\\
D_\bw q_w&=0,\\
D_\bw \tq^\dag&=0,\\
D_\bw\phi_z&=0.
\end{align*}
The last two of these equations imply that generically
$\tq^\dag=\phi_z=0$. The first two equations are called the
nonabelian vortex equations. They generalize the Hitchin equations
to the case when the Higgs field is in representation $R$ of $G$
other than the adjoint. We will denote the moduli space of solutions
$\MV(G,R,\Sigma)$. Since $q_w, q_\bw$ are bosonic coordinates on
$\MV$, it is natural to define the ghost number so that both of
these fields have $gh=0$. As discussed in section \ref{sec:BRST},
this requirement fixes uniquely the ghost numbers of all matter
fields.

The moduli space $\MV$ is well-known to be K\"ahler, with the
holomorphic coordinates being $A_\bw$ and $q_w$, and symplectic form
$$
\omega_V=\frac{i}{2\pi}\int_\Sigma d^2w \left(\Tr\,\delta
A_\bw\wedge \delta A_w+\delta q_w \wedge \delta q_\bw\right).
$$
We denote the corresponding complex structure $I_V$. The K\"ahler
form of the sigma-model is $\Im\tau\,\omega_V$.

Another way to describe the complex structure on $\MV$ is to
consider the space of pairs $(E,\frq)$, where $E$ is a holomorphic
$G$-bundle on $\Sigma$ and $\frq=q_w dw$ is a holomorphic section of
the vector bundle $R(E)\otimes K_\Sigma$. We will call such a pair a
$(G,R)$ Higgs bundle on $\Sigma$. Clearly, any solution of the
nonabelian vortex equations defines a $(G,R)$ Higgs bundle.

Similar objects have been studied by mathematicians under the name
``holomorphic pairs'' \cite{B,BD,HL,BDW}. A holomorphic pair on
$\Sigma$ is a holomorphic vector bundle $E$ together with a section
of $E\otimes V$, where $V$ is a fixed vector space. A $(G,R)$ Higgs
bundle is a modification of this: instead of a vector bundle we have
a principal $G$-bundle, and instead of $E\otimes V$ we have a vector
bundle $R(E)\otimes K_\Sigma$.\footnote{The notion of a
``holomorphic pair'' as defined in \cite{HL} is very general and
includes holomorphic pairs of \cite{B} and our $(G,R)$ Higgs bundles
as special cases. See also \cite{BDGW} for a survey of various kinds
of vortex equations and associated holomorphic data.}

In the theory of holomorphic pairs and related objects, an important
role is played by the notion of (semi)stability. One can define the
moduli space of semistable holomorphic pairs and show that it is
isomorphic to the moduli space of ``vortex equations'' very similar
to our vortex equations \cite{B,BD,BDW}. Similarly, one can
formulate a notion of (semi)stability for $(G,R)$ Higgs bundles.
(For a discussion of stability for holomorphic pairs and related
objects, see \cite{BDGW}.) For simplicity, we will spell it out only
for $G=SU(2)$ and when $R$ is the sum of $k$ copies of the
fundamental representation. Such a $(G,R)$ Higgs bundle is called
stable (resp. semistable) if any holomorphic line subbundle $F$ of
$E$ satisfying $\frq\in H^0(F\otimes K_\Sigma)\otimes \CC^k$ has
negative (resp. nonpositive) first Chern class. One can easily check
that for this special choice of $G$ and $R$ any solution of vortex
equations defines a semistable $(G,R)$ Higgs bundle. Presumably, the
moduli space of solutions of nonabelian vortex equations is
isomorphic to the moduli space of semistable $(G,R)$ Higgs
bundles.\footnote{A special case, corresponding to the twisted
$\cN=4$ gauge theory, is when $R$ is the adjoint representation. In
this case a $(G,R)$ Higgs bundle is the same as a Higgs bundle, as
defined by Hitchin \cite{Hitchin}, and the vortex equations become
the Hitchin equations. The isomorphism of the moduli space of
Hitchin equations and the moduli space of semistable Higgs bundles
is known to hold for arbitrary $G$ \cite{Hitchin}.}

We know from the previous section that for $t\neq 0,\infty$ the
twisted gauge theory depends on the complex structure on $\Sigma$
but not on the gauge coupling $e^2$. Since the complex structure on
$\MV(G,R,\Sigma)$ depends on the complex structure on $\Sigma$,
while $e^2$ sets the overall scale of the K\"ahler form, it is clear
that the topological sigma-model to which the gauge theory reduces
must be the B-model with target $\MV(G,R,\Sigma)$. We can check that
the theory on $C$ is indeed the B-model by looking at the equations
defining BRST-invariant bosonic configurations. We find:
\begin{align*}
F_{\bw z}&=0,\\
F_{\bw \bz}&=0,\\
D_z q_w&=0,\\
D_\bz q_w&=0.
\end{align*}
These equations are overdetermined and require the map from $C$ to
$\MV$ to be both holomorphic and anti-holomorphic. This is precisely
what one expects for the B-model on $C$ with target
$\MV(G,R,\Sigma)$.

The two-dimensional interpretation of the theta-angle is the same as
in \cite{KW}. {}From the viewpoint of the effective field theory on
$\Sigma$ it produces a flat but topologically nontrivial B-field on
$\MH(G,C)$ of the form
$$-\omega_I \Re\,\tau\ ,
$$
where $I=J_\infty$. {}From the viewpoint of the effective field theory
on $C$, it produces a flat B-field on $\MV(G,R,\Sigma)$ in the same
cohomology class as
$$
-\omega_V \Re\,\tau\ .
$$

\subsection{Some properties of the reduced theories}

Now we can explain some of the results of section \ref{action} in
two-dimensional terms. We showed there that for $t\neq 0,\infty$ the
$\beta''$-model does not depend on the parameter $\tau$. {}From the
viewpoint of the half-twisted model on $\Sigma$, this happens
because for $t\neq 0,\infty$ the form $\omega_t$ is exact, and
therefore are no nontrivial worldsheet instantons. {}From the
viewpoint of the B-model on $C$, there is no dependence on $\Im\tau$
because it affects only the K\"ahler form of $\MV$, and there is no
$\Re\,\tau$ dependence because the B-field is of type $(1,1)$.

We also showed that the twisted gauge theory is topological along
$C$ and holomorphic along $\Sigma$. This is also obvious from the
two-dimensional viewpoint, since the B-model on $C$ is a 2d TFT,
while the half-twisted model on $\Sigma$ is a chiral CFT.

Let us also comment on the two special cases $t=0$ and $t=\infty$.
{}From the viewpoint of the half-twisted field theory on $\Sigma$,
these points are special because the forms $\omega_0=-\omega_I$ and
$\omega_\infty=\omega_I$ are not exact, and therefore worldsheet
instanton contributions introduce nontrivial dependence on $\tau$.
More precisely, since instanton contributions depend holomorphically
on the combination $B+i\omega,$ the correlators at $t=\infty$ (resp.
$t=0$) are antiholomorphic (resp. holomorphic) functions of $\tau$.
{}From the viewpoint of the effective field theory on $C$, for
$t=0,\infty$ the model degenerates into a half-twisted $(2,2)$ model
with target $\MV(G,R,\Sigma)$ and complex structure $\pm I_V$. One
can see it, for example, from the fact that the BPS equations become
elliptic rather than over-determined and require the map to $\MV$ to
be holomorphic. In such a model correlators depend holomorphically
on $B+i\omega$, which again translates into holomorphic or
anti-holomorphic dependence on $\tau$ depending on whether $t=0$ or
$t=\infty$.

\subsection{Anomaly cancelation}

To define the $\beta''$-model, it was essential that the $U(1)_N$
symmetry is anomaly-free, that is, that the condition (\ref{ac}) is
satisfied. We can also explain the significance of this condition
from the two-dimensional viewpoint.

{}From the viewpoint of the B-model on $C$, this is simply the
Calabi-Yau condition on the target $\MV(G,R,C)$. Indeed, the
$U(1)_N$ symmetry is nothing but the R-symmetry of the model, which
is nonanomalous precisely when the target manifold is a Calabi-Yau.

If we pretend that $\MV(G,R,\Sigma)$ is smooth, we can see it more
explicitly as follows. After identifying $\MV$ with the moduli space
of stable $R$-Higgs bundles, we can describe a holomorphic tangent
vector at a point $(E,0)\in\MV$\footnote{The moduli space $\MV$ is
homotopic to its submanifold $q_w=0$, therefore it is sufficient to
compute the restriction of the first Chern class to this
submanifold.} as the equivalence class of a pair $(\delta
A_\bw,\delta q_w)$ where $\delta A_\bw$ is an arbitrary variation of
$A_\bw$ and $\delta q_w$ satisfies
$$
D_\bw\delta q_w=0.
$$
The equivalence relation is
$$
\left(\delta A_\bw,\delta q_w\right)\sim \left(\delta A_\bw+D_\bw a,
\delta q_w\right),
$$
where $a$ is a section of $\ad (E)$. Thus the tangent space at a
point $(E,0)$ can be identified with
$$
H^1(\ad(E))\oplus H^0(R(E)).
$$
If $\MV$ were smooth, the first Chern class of this bundle could be
equated with the first Chern class of the virtual tangent bundle.
The latter can be computed via the family index theorem, giving
\begin{multline}\label{anomalyone}
c_1(T\MV^{virt})=-\int_\Sigma
\left(\ch_4(\ad(\cE_R))-\ch_4(R(\cE_R))\right)\\
= -\left(1-\frac{C(R)}{C(G)}\right)\int_\Sigma \ch_4(\ad(\cE_R)).
\end{multline}
where $\cE_R$ is the universal $G$-bundle on $\MV(G,R,\Sigma)\times
\Sigma$ and $\ch_4$ is the degree-4 part of the Chern character. The
anomaly cancelation condition (\ref{ac}) ensures that this vanishes.
In fact, $\MV(G,R,\Sigma)$ is not smooth, and it is reasonable to
{\it define} its first Chern class as the $U(1)_N$ anomaly. The
latter is still given by (\ref{anomalyone}).

{}From the viewpoint of the half-twisted model on $\Sigma$, the
interpretation of the condition (\ref{ac}) is rather  different. In
general, for a $(4,0)$ sigma-model with target $X$ and fermionic
bundle $\cR$ to be well-defined (even before twisting), the degree-4
parts of the Chern characters of $\cR$ and $X$ must be equal:
\begin{equation}\label{anomalytwo}
\ch_4(X)=\ch_4(\cR).
\end{equation}
This condition is necessary for the Pfaffian of the right-handed
Dirac operator to be well-defined. We claim that in our case this
condition is equivalent to (\ref{ac}). Indeed, the bundle $\cR$ in
our case is equal to the fiberwise cohomology of the (twisted)
bundle $R(\cE)$ with respect to the projection $\MH(G,C)\times C\ra
\MH(G,C)$. The families index theorem gives the following virtual
Chern character for $\cR$ in degree four:
$$
\ch_4(\cR)=(g-1)\ch_4(R(\cE))_\MH,
$$
where subscript $\MH$ indicates that we must take the piece of the
cohomology class which is pulled back from $\MH$. The tangent bundle
to $\MH$ is similarly the fiberwise cohomology of $\ad(\cE)$, so we
get
$$
\ch_4(\MH)=(g-1)\ch_4(\ad(\cE))_\MH.
$$
The equality (\ref{anomalytwo}) holds precisely if $C(G)=C(R)$.

\subsection{Chiral algebras from gauge theories in four dimensions}

We have seen that to any $\cN=2$ superconformal gauge theory one can
attach a pair of chiral 2d CFTs, one being a half-twisted $(4,0)$
sigma-model with target $\MH$ and the other being a half-twisted
$(2,2)$ sigma-model with target $\MV$. These chiral algebras encode
certain holomorphic sectors of the four-dimensional gauge theory and
generalize the familiar chiral ring.

One can generalize this construction in two directions. First, one
can start with an $\cN=1$ supersymmetric gauge theory with a
nonanomalous $U(1)_R$ symmetry and twist it suitably along $C$ and
$\Sigma$. This gives a chiral CFT on $C$ or $\Sigma$ which is a
half-twisted $(2,0)$ sigma-model with target $\MV$. Such twisted
$\cN=1$ gauge theories have been previously considered by A.
Johansen \cite{Johansen}.

As an example, let us consider $\cN=1$ super-QCD, i.e. an $\cN=1$
gauge theory with $G=SU(N)$ and $2k$ chiral superfields $Q_i,\tQ_i,
i=1,\ldots,k,$ where the superfield $Q_i$ takes values in the
fundamental representation of $SU(N)$ and $\tQ_i$ takes values in
its dual. This theory is not superconformal: for $k<3N$ it is
asymptotically free and flows to strong coupling in the infrared,
while for $k\geq 3N$ it is free in the infrared and has a Landau
pole.

A nonanomalous R-symmetry exists such that the lowest components of
both $Q_i$ and $\tQ_i$ have R-charge $1-N/k$ \cite{Seiberg}. The
R-current is unique up to addition of a multiple of the $U(1)_B$
current ($Q_i$ and $\tQ_i$ have $U(1)_B$ charge $1$ and $-1$
respectively). We consider the theory on $C\times\Sigma$ and shift
$U(1)_C$ charge by half the R-charge, so that the left-handed
gaugino becomes a section of
$$
K_\Sigma^{-1/2}\otimes K_C + K_\Sigma^{1/2}\otimes \cO_C,
$$
while the right-handed gaugino becomes a section of
$$
K_\Sigma^{-1/2}\otimes K_C^{-1}+K_\Sigma^{1/2}\otimes \cO_C.
$$
We see that the theory twisted along $C$ has two supercharges of the
same chirality which are Hermitian-conjugate of each other. That is,
the effective field theory on $\Sigma$ has $(2,0)$ supersymmetry. We
are still free to add a multiple of the $U(1)_B$ current to the
R-current; one may choose it so that the lowest component of $Q_i$
becomes a section of $K_C$ (for all $i$); the lowest component of
$\tQ_i$ will then be a section of $K_C^{1-2N/k}$. Note that $\tQ_i$
field has a fractional $U(1)_C$ charge, in general; in order to have
only integer spins in the twisted theory one is forced to restrict
the values of $N$ and $k$ suitably. The most natural choice is
$k=2N$; then the reduced theory on $\Sigma$ is similar to what we
had in the $\cN=2$ case. Namely, it is a $(2,0)$ sigma-model whose
target is $\MV(SU(N),R,C)$, where $R$ is the sum of $2N$ copies of
the fundamental representation of $SU(N)$.

This theory still depends on the K\"ahler form of $C$, so the
reduced theory becomes equivalent to the gauge theory on
$C\times\Sigma$ only when $C$ shrinks to zero size. We can try to
rectify this by twisting it along $\Sigma$ with a suitable R-current
(which is again a linear combination of the canonical R-current
described above and the $U(1)_B$ current). In this way one obtains a
half-twisted $(2,0)$ sigma-model with target $\MV$. Correlators in
this theory depend holomorphically on coordinates on $\Sigma$.

Alternatively, one can start with a finite $\cN=2$ gauge theory in
four dimensions but allow $C$ to have a boundary, with suitable
BRST-invariant boundary conditions, thus halving the number of
supercharges. {}From the viewpoint of the effective theory on $C$,
this corresponds to a B-brane on $\MV$. {}From the viewpoint of the
field theory on $\Sigma$, we get a half-twisted $(2,0)$ sigma-model
on $\Sigma$ which depends on the choice of the brane. In particular,
to any B-brane on $\MH(G,\Sigma)$ (in complex structure $I$ in the
notation of \cite{KW}) one can associate a chiral algebra on
$\Sigma$.

\section{Observables of the twisted theory}\label{observables}

\subsection{Local observables and their descendants}
First, let us assume that $t\neq 0,\infty$. By inspection, the only
observables in the gauge theory which are BRST invariant, not
BRST-exact, gauge-invariant, and local are gauge-invariant
polynomials made of $q_w$ and $\tq^\dag.$ Note that covariant
derivatives of these operators along $\bw$ are BRST-exact; this
implies that all correlators and OPE coefficients of gauge-invariant
polynomials are holomorphic functions of $w$. Similarly, the
covariant exterior differential $\cD_C$ applied to these operators
gives BRST-exact operators; this implies that correlators and OPE
coefficients are independent of $z$. This confirms our claim that
the theory is topological along $C$ and holomorphic along $\Sigma$.

The OPE of these local observables is nonsingular, because of
Hartogs' theorem: a holomorphic function, such as an OPE coefficient
function, cannot have a singularity in complex codimension 2. An
alternative way to show this is to note that because of topological
invariance along $C$, when computing the singularities of the OPE
coefficient functions we can keep the insertion points on $C$ far
from each other. But then the coefficient functions cannot becomes
singular even if the insertion points on $\Sigma$ collide. Since the
OPE coefficients are nonsingular and holomorphic, they must be
constant. Therefore local observables form a commutative algebra.
This algebra is a subalgebra of the anti-chiral ring of the
untwisted theory.

For example, let $G=SU(N)$ and let $R$ be the sum of $2N$ copies of
the fundamental representation. The theory has $SU(2N)$ global
flavor symmetry which acts only on the matter fields; accordingly,
we will denote the fields as $\tq^\dag_i$ and $q_w^i$, where
$i=1,\ldots,2N$ is the $SU(2N)$ flavor index. Then the algebra of
gauge-invariant polynomials is generated by the ``mesons''
$$
M^i_j=q_w^i\tq^\dag_j,
$$
the ``baryons''
$$
{\tilde B}_{i_1\ldots i_N}=\tq^\dag_{i_1}\ldots \tq^\dag_{i_N}
$$
and the ``anti-baryons''
$$
B^{i_1\cdots i_N}=q_w^{i_1}\ldots q_w^{i_N}.
$$
Here we did not show explicitly how the $SU(N)$ color indices are
contracted, since there is a unique way to do so.

We can construct nonlocal BRST-invariant and gauge-invariant
observables following the descent procedure introduced by E. Witten
in the context of topological field theory \cite{WittenTFT}. In a
TFT, this procedure allows one to construct ``pre-observables'':
differential forms whose BRST-variation is exact. One can obtain
BRST-invariant observables by integrating pre-observables over
homology cycles.

The $\beta''$-model is topological along $C$ and holomorphic along
$\Sigma$. The descent procedure along $C$ works in the usual manner,
giving observables which are integrals over 1-cycles and the
fundamental class of $C$. On the other hand, descent along $\Sigma$
gives pre-observables which are $(0,1)$ forms on $\Sigma$ whose
BRST-variations are $\bpartial_\Sigma$-exact. We can get true
observables by multiplying these by holomorphic 1-forms on $\Sigma$
and integrating over $\Sigma$.

To be concrete, let us again consider the case $G=SU(N)$ and let $R$
be the sum of $2N$ copies of the fundamental representation. The
first descendant of the anti-baryon $B=q_w^N$ along $C$ is defined
by
$$
\delta_t B^{(C,1)}=d_C B.
$$
This equation is solved by
$$
B^{(C,1)}=-\frac{Nt^{-1}}{4} \Psi_w q_w^{N-1},
$$
where the $SU(N)$ indices are completely anti-symmetrized. This
pre-observable is a section of $K_\Sigma^N\times \Omega^1_C$ and can
be integrated over a 1-cycle $\gamma\in H_1(C)$ to give a true
observable
$$
B^{(C,1)}(\gamma,p)=\int_{\gamma\times p} B^{(C,1)}, \quad
p\in\Sigma
$$
Such observables are local on $\Sigma$ and may have nontrivial OPEs
with each other. OPE is necessarily holomorphic in the coordinate
$w$. Another constraint on the OPE comes from the fact that the
theory is topological along $C$: if the homology cycles $\gamma$ and
$\gamma'$ have vanishing intersection number, one can choose their
representatives to be nonintersecting on $C$, and then there can be
no singularity in the OPE on $\Sigma$. Similarly, diffeomorphism
invariance along $C$ guarantees that the OPE of $B^{(C,1)}$ with any
local observable on $C\times\Sigma$ is nonsingular.

Now let us consider descent along $\Sigma$. Taking the same
observable $B$, its first (and last) $\Sigma$-descendant is defined
by
$$
\delta_t B^{(\Sigma,1)}_{\bw}=\partial_\bw B.
$$
We find
$$
B^{(\Sigma,1)}_\bw=-\frac{N t^{-1}}{8}\eta_{w\bw}q_w^{N-1},
$$
where $\eta_{w\bw}$ was defined in (\ref{eta}). This pre-observable
is a $(0,1)$ form on $\Sigma$ with values in $K_\Sigma^N$. To
construct from it a true observable, we must integrate over $\Sigma$
a product of $B^{(\Sigma,1)}$ and a holomorphic section of
$K_\Sigma^{1-N}$. In other words, one can view $B^{(\Sigma,1)}$ as
an observable taking values in the vector space $H^1(K_\Sigma^N)$
which by Serre duality is dual to $H^0(K_\Sigma^{1-N})$. These
observables are local on $C$; because of diffeomorphism invariance
along $C$ their correlators do not depend on the $z$-coordinate
(i.e. the OPE is nonsingular).

Finally let us discuss the case $t=\infty$ (the case $t=0$ is very
similar). The adjoint Higgs field $\phi_z$ is now BRST-closed, while
$\lambda_z$ is no longer BRST-exact. We also note that for
$t=\infty$ the covariant derivatives $D_\bw\phi_z, D_\bz\phi_z,$ and
$D_\bw\lambda_z, D_\bz\lambda_z$ are BRST-exact. Thus we can build
new local observables as gauge-invariant polynomials built out of
$\phi_z$ and $\lambda_z,$ and their correlators will be holomorphic
functions of both $z$ and $w$. Note also that $\phi_z$ and
$\lambda_z$ commute up to BRST-exact terms:
$$
[\phi_z,\lambda_z]\sim \delta_\infty D_z\phi_z.
$$
Thus we get BRST-invariant and gauge-invariant local observables
$$
\Tr\, \phi_z^m \lambda_z^l.
$$
Because of Hartogs' theorem, their OPE coefficients are nonsingular,
so these operators generate a supercommutative algebra.

Of course, we still have the observables such as $B,\tB,$ and $M$,
which are part of the anti-chiral ring.\footnote{Note that certain
other candidate observables which mix matter fields and fields in
the gauge multiplet, such as $ q_w\phi_z^m \lambda_z^l \tq^\dag,$
are BRST-exact.} It may seem surprising that OPE of operators such
as $\Tr\, \phi_z^k$ on one hand and $M,\tB,B$ on the other hand are
nonsingular, since the former belong to the chiral ring of the
untwisted theory while the latter belong to the anti-chiral ring.
But this becomes quite obvious once we apply an $SU(2)_R$
transformation which in the untwisted theory maps $q_w$ to $\tq$ and
$\tq^\dag$ to $-q_\bw$ and leaves $\phi_z$ invariant. After this
transformation, all the local operators found above become part of
the usual chiral ring, and the regularity of the OPE coefficients
follows in the usual manner. This is true even when we take into
account the local observables containing $\lambda_z$: after the
above-mentioned $SU(2)_R$ rotation, $\lambda_z$ becomes the lowest
component of the chiral superfield
$W_\alpha=-\frac{1}{8}\bD^2e^{-2V}D_\alpha e^{2V}$.

For $t=\infty$, the theory is a chiral CFT both along $C$ and
$\Sigma$. Thus the descent procedure applied to the above operators
gives us nonlocal observables valued in $H^1(C\times\Sigma,\cL)$ and
$H^2(C\times\Sigma,\cL)$ for various holomorphic line bundles $\cL$
on $C\times\Sigma$. These observables can have nontrivial
holomorphic OPE with local observables and between themselves. By
way of an example, consider the observable $\Tr\, \lambda_z\phi_z$.
It descendant along $C$ is
$$
\Tr \,
\left(F_{z\bz}\phi_z+\frac{1}{4}\lambda_z\lambda_{z\bz}\right),
$$
while its descendant along $\Sigma$ is
$$
\Tr \, \left(-F_{\bw z}\phi_z+\frac{1}{4}\lambda_z\lambda_{\bw
z}\right).
$$
We can combine them into a $(0,1)$ form on $C\times\Sigma$ with
values in $K_C^2$; this defines an observable valued in
$H^1(K_C^2)$.

\subsection{Wilson-'t Hooft loop observables}\label{loopops}

For $t\neq 0,\infty$ the twisted theory has another interesting
class of nonlocal gauge-invariant observables. The simplest of these
are gauge-invariant functions of the holonomy of the BRST-invariant
connection
$$
\cA_C=\cA_z dz+\cA_\bz d\bz.
$$
Explicitly, for any closed curve $\gamma$ on $C$, any point $p$ on
$\Sigma$, and any irreducible representation $R$ of the gauge group
$G$ we can define an observable
$$
W(R,\gamma,p)=\Tr_R P\exp\left(i\int_{\gamma\times p}\cA\right)
$$
We will call it the Wilson loop observable. It depends only on the
homotopy class of $\gamma$ (unlike superficially similar observables
constructed in the previous section, which are labelled by homology
classes of $C$). Unlike the case of the GL-twisted theory, such
observables exist for all $t\neq 0,\infty$. Since an irreducible
representation is completely determined by its highest weight,
sometimes we will label Wilson loops by elements of the weight
lattice $\Lambda_w(G)$ modulo the action of the Weyl group.

The Wilson loop observables are local observables in the effective
field theory on $\Sigma$. {}From the viewpoint of the twisted $(4,0)$
model with target $\MH$, they generate the algebra of holomorphic
functions on $\MH$. {}From the 2d viewpoint, it is clear that the OPE
of Wilson loops is nonsingular, irrespective of the choice of
$\gamma$; one can also show this directly in the gauge theory, see
below.

Analogy with $\cN=4$ gauge theory suggests that there should also be
BRST-invariant 't Hooft loop operators, which are magnetic analogs
of Wilson loops. In fact, S-duality requires the existence of such
operators. As in \cite{K,KW}, we can construct them by requiring the
fields in the path integral to have a prescribed singularity along
$\gamma\times p$. The singularity should be compatible with BRST
transformations, in the sense that there should exist bosonic
configurations which are BRST-invariant and have the required
singular behavior near $\gamma\times p$.

To write down the singularity in the fields corresponding to an 't
Hooft operator, we may replace an arbitrary curve $\gamma\subset C$
by a straight line on $\CC$ oriented in the direction $x^1=\Re\, z$.
The gauge field will be required to have a monopole singularity at
$x^2=x^3=x^4=0$:
$$
F\sim \frac{\mu}{2}\star_3 d\frac{1}{r}.
$$
Here $r^2=(x^2)^2+(x^3)^2+(x^4)^2$, $\star_3$ is the Hodge star
operator in the $234$ plane, and $\mu\in\g$ is the so-called GNO
charge of the monopole. It is the value at $1$ of a homomorphism
from $\RR$ to $\g$ obtained from a homomorphism of Lie groups
$U(1)\ra G$. The GNO charge is defined modulo the adjoint action of
$G$, so one can assume that it belongs to a chosen Cartan subalgebra
of $\g$. Then $\mu$ becomes a coweight of $G$ defined modulo the
action of the Weyl group $\cW$. We will denote by $T_\mu(\gamma,p)$
the 't Hooft operator associated with a coweight $\mu$.

The singularity in the Higgs field $\phi$ is uniquely fixed by the
requirement of BRST-invariance:
$$
\phi_z\sim -\frac{t^{-1}\mu}{4r},\quad \phi_\bz\sim-\frac{t\mu}{4r}.
$$
(Note that this is compatible with the reality condition
$\phi_z^\dag=\phi_\bz$ only if $|t|=1$. This does not mean that 't
Hooft operators make sense only if $|t|=1$: when evaluating the
path-integral using saddle-point approximation it is quite common to
expand about complex solutions to saddle-point equations.) Thus for
any $t\neq 0,\infty$ we have 't Hooft operators labelled by
equivalence classes of coweights of $G$, or, equivalently, by the
equivalences classes of weights of the Langlands dual group $\LG$.
This suggests that there could be a dual description of the theory
where the gauge group is $\LG$. This is believed to be true in the
special case when the hypermultiplet is in the adjoint
representation of $G$; the dual theory is conjectured to be
identical to the original one except $G$ is replaced with $\LG$. For
other theories, such as $G=SU(2)$ and hypermultiplets in the
defining representation, the naive duality conjecture appears to be
false,\footnote{The putative dual description of the theory with
$G=SU(2)$ and four hypermultiplets in the defining representation,
would have to have dynamical magnetic sources whose magnetic charge
is half the charge of the 't Hooft-Polyakov monopole; this seems
impossible in the context of semiclassical gauge theory.} but a
somewhat different version of S-duality appears to hold, see section
\ref{Sduality} below.

We see that in the $\beta''$-model one can construct Wilson and 't
Hooft operators which commute with the same BRST operator. This is
contrast to the GL-twisted theory, where for a fixed BRST operator
we have either Wilson observables or 't Hooft observables, or
neither, but never both. Moreover, in the $\beta''$-model by fusing
Wilson and 't Hooft operators we can get more general Wilson-'t
Hooft operators labelled \cite{K} by equivalence classes of pairs
$$
(\mu,\nu),\quad \mu\in \Lambda_{cw}(G),\ \nu\in \Lambda_w(G)
$$
under the action of the Weyl group as well as by elements of
$\pi_1(C)$.

All Wilson-'t Hooft operators are local on $\Sigma$ and in general
their correlators are holomorphic functions of the coordinates on
$\Sigma$. In particular, in general there is a nontrivial OPE
between different loop operators, as we will see below. But if we
choose a particular element $y\in \pi_1(C)$ and only consider loop
operators corresponding to this homotopy class, then the OPE is
nonsingular. Indeed, because of diffeomorphism invariance along $C$,
we can choose two loops $\gamma$ and $\gamma'$ representing the
class $y$ which do not intersect on $C$. Given two loop operators
$A$ and $B$ corresponding to the class $y$, we may choose $A$ and
$B$ to be supported on $\gamma\times p$ and $\gamma'\times p'$,
where $p,p'\in\Sigma$. Then the operator product will remain
nonsingular as $p$ merges with $p'$, because the supports of $A$ and
$B$ on $C\times\Sigma$ remain disjoint in this limit.

We conclude that loop operators corresponding to a fixed
$y\in\pi_1(C)$ form a commutative algebra, and that there is a
natural basis in this algebra labelled by
$\left(\Lambda_{cw}\times\Lambda_w\right)/\cW$. It is an interesting
problem to determine the structure constants of the algebra in this
basis, i.e. determine the ``fusion rules'' of Wilson-'t Hooft
operators. We are planning to discuss this issue elsewhere; here we
simply note that the answer will depend not only on the group $G$
but also on the representation $R$ of the hypermultiplet. This is
clear from the fact that the algebra structure should be compatible
with the action of S-duality, which works differently for different
$R$. In the case when $R$ is the adjoint representation, the algebra
seems to be abstractly isomorphic to the $\cW$-invariant part of the
group algebra of $\Lambda_{cw}\times\Lambda_w$. The latter has an
obvious basis labelled by the correct set: one simply takes an
element $(\mu,\nu)\in \Lambda_{cw}\times\Lambda_w$ and averages over
the Weyl group. Nevertheless, it is not the basis we want. For
example, if we take the naive basis elements of the form $(0,\nu_1)$
and $(0,\nu_2)$, their product will have the form
$$
\sum_{x\in\cW}(0,\nu_1+x\cdot\nu_2),
$$
where $x\cdot\nu$ denotes the result of applying an element
$x\in\cW$ to $\nu\in\Lambda_w$. On the other hand, the expected
answer (the operator product of Wilson loops labelled by $\nu_1$ and
$\nu_2$) is given by the decomposition of the tensor product of
representations of $G$ with highest weights $\nu_1$ and $\nu_2$ into
irreducibles.

Let us now briefly discuss the OPE of Wilson loop operators which
wrap different cycles on $C$. We claim that for such operators there
is no singularity in the OPE. Indeed, if we compute the OPE in
perturbation theory using Wick theorem, the singularity in the OPE
may come only from the propagators of $A$ and $\phi_z$. Since the
propagators are proportional to $e^2$, and we know that varying
$e^2$ changes the action only by BRST-exact terms, we conclude that
the OPE must be nonsingular.

We conclude this section by displaying an example where the OPE of
loop operators is nontrivial. Consider the case $G=U(1)$. The
abelian $\cN=4$ gauge theory is free, and the OPE of arbitrary loop
observables can be readily computed. As argued above, the OPE of
arbitrary Wilson operators is trivial, and by S-duality the same
should be true for 't Hooft operators. But as we will now explain,
OPE of a Wilson operator and an 't Hooft operator is nontrivial, in
general. It is particularly simple to see this from the viewpoint of
the effective field theory on $\Sigma$. Then we are dealing with a
twisted $(4,0)$ sigma-model whose target is the moduli space of flat
$\CC^*$ connections on $C$. As a complex manifold, it is isomorphic
to $(\CC^*)^{2g}\simeq T^*(S^1)^{2g}$, and therefore the sigma-model
has the usual momentum and winding states. The winding charge takes
values in the lattice $H_1(C,\ZZ)$; the momentum also takes values
in a lattice which is most naturally identified with $H^1(C,\ZZ)$,
but for our purposes it is more convenient to make use of the
Poincar\'{e} isomorphism $H^1(C,\ZZ)\simeq H_1(C,\ZZ)$ and regard
momentum as taking values in $H_1(C,\ZZ)$. There is a natural
pairing between winding and momentum lattices which with our
conventions can be identified with the intersection pairing; we will
denote it $\langle m, n\rangle$, where $m,n\in H_1(C,\ZZ)$ are
winding and momentum respectively. As noted in \cite{HMS}, the
Wilson loop $W_n(\gamma)$ upon Kaluza-Klein reduction corresponds to
a conformal primary with momentum $n[\gamma]$, where $[\gamma]$ is
the homology class of $\gamma$. Similarly, the 't Hooft operator
$T_m(\gamma)$ carries winding $m[\gamma]$. BRST-invariance requires
local operators on $\Sigma$ corresponding to Wilson and 't Hooft
loop operators also to carry imaginary momentum in the ``cotangent''
directions on $T^*(S^1)^{2g}$. To infer the singularity in the OPE,
we recall that in the presence of a winding-state operator of charge
$m$ an operator with momentum $n$ has a monodromy $2\pi \langle
m,n\rangle$. By holomorphy on $\Sigma$, this implies that the OPE
has a singularity of the form $w^{\langle m,n\rangle}$, where $w$ is
the insertion point of the Wilson operator relative to the 't Hooft
operator.

\section{The action of loop operators on branes}\label{branes}

In \cite{KW}, loop operators in topologically twisted gauge theory
have been connected to geometric Langlands program via their action
on branes. The fact that loop operators act by functors on the
category of topological D-branes is also very useful for computing
the Operator Product Expansion of loop operators. Our goal in this
section is to generalize some of these considerations to the
$\beta''$-model.

Depending on whether $C$ or $\Sigma$ is taken to be small, the
twisted gauge theory reduces to either a half-twisted $(4,0)$
sigma-model on $\Sigma$, or a B-twisted $(2,2)$ sigma-model on $C$.
The half-twisted sigma-model on $\Sigma$ is chiral, and therefore
$\Sigma$ cannot have a nonempty boundary. But it is perfectly
possible to allow $C$ to have a boundary. A choice of BRST-invariant
boundary conditions on $C$ defines a topological D-brane of type B
on the K\"ahler manifold $\MV$.

All this matches very well with the spectrum of BRST-invariant loop
operators. We have seen that there exist Wilson-'t Hooft loop
operators which are extended on $C$ and point-like on $\Sigma$. If
one lets such a loop operator run along the boundary of $C$, one
gets a functor from the category of B-branes on $\MV$ to itself. All
such functors necessarily commute. {}From the point of view of the
half-twisted sigma-model on $\Sigma$, loop operators are simply
local BRST-invariant operators.

The action of Wilson operators on the category of branes is easy to
determine. The same arguments as in \cite{KW} tell us that a Wilson
operator in representation $\cS$ of the group $G$ tensors the brane
with the holomorphic vector bundle $\cS(\cE_p)$ where $\cE$ is the
universal $G$-bundle on $\MV\times \Sigma$, $p\in\Sigma$ is the
insertion point of the Wilson operator, and $\cE_p$ is the
restriction of $\cE$ to $\MV\times \{p\}$.

It is more difficult to understand the action of 't Hooft operators.
The first step is to consider the BPS equations in the case when
$C=\RR\times I$, where $I$ is an interval, and the fields are
assumed to be time-independent. For definiteness, let $t=i$ (as
discussed above, nothing depends on the choice of $t$, provided
$t\neq 0,\infty$). Let $x^1$ parameterize the ``time'' direction,
and $x^2$ be the coordinate on $I$. We will assume that the 't Hooft
operator is inserted at $x^2=0$. In the gauge $A_2=0$ the static BPS
equations have the form
\begin{align*}
\partial_2 A_\bw&=-2D_\bw \phi_z,\\
\partial_2 q_w&=-2iq_w\phi_z.
\end{align*}
One can regard these equations as describing the evolution of the
$(G,R)$-Higgs bundle $(E,\frq)$ on $\Sigma$ as one varies $x^2$. The
equations say that $(E,\frq)$ changes only by a complex gauge
transformation until on gets to $x^2=0$ where $\phi_z$ has a
singularity. There the $G$-bundle $E$ undergoes a Hecke
transformation, the same one as in the twisted $\cN=4$
theory~\cite{KW}. We will denote by $E_-$ and $E_+$ the bundles for
$x^2<0$ and $x^2>0$. We remind that the Hecke transformation
modifies the bundle at one point $p\in \Sigma$, i.e. one is given a
holomorphic isomorphism between  $E_+$ and $E_-$ on
$\Sigma\backslash p$. But a section of $E_-$ which is holomorphic at
$p$ may become meromorphic when regarded as a section of $E_+$.
Therefore the requirement that the Higgs field $q_w$ is holomorphic
both for $x^2<0$ and $x^2>0$ puts a constraint on the allowed Hecke
transformations.

As an illustration, consider $G=SU(2)$, and take $R$ to be the sum
of four copies of the fundamental representation. Consider an 't
Hooft operator with
$$
\mu=\begin{pmatrix} 1 & 0\\ 0 & -1\end{pmatrix}
$$
inserted at the point $w=0$ on $\Sigma$. The Higgs field $\frq$ is a
made of four holomorphic sections of the bundle $E_-\times
K_\Sigma$. Since we are working locally on $\Sigma$, we may assume
that $E_-$ and $K_\Sigma$ are trivial, so we may think of these
sections as sections of $E_-$.

Let $V$ be the subspace of $E_-\vert_{w=0}$ spanned by the four
sections at $w=0$. If $V$ is zero-dimensional (i.e. if all four
sections vanish at $w=0$), then the 't Hooft operator can effect an
arbitrary Hecke transformation on the bundle $E_-$; the space of
such transformations is isomorphic to $T\PP^1$. Let us recall how
this comes about \cite{KW}. A Hecke transformation is specified once
we pick a pair of holomorphic sections of $E_-$ which generate $E_-$
in the neighborhood of $w=0$. If these sections are $s_1$ and $s_2$,
then $E_+$ is generated by
$$
s'_1=w^{-1} s_1,\quad s'_2 =w s_2.
$$
Not every choice of $s_1,s_2$ gives a different $E_+$, since any
holomorphic $SL(2)$ gauge transformation of $s'_1,s_2'$ with
holomorphic coefficients gives the same $E_+$. Using this freedom,
one can show that only the first two terms in the Taylor expansion
of $s_1$ affect $E_+$; moreover, if $s_1(w)$ has the form
$$
s_1(w)=\begin{pmatrix} a_1+b_1 w+\ldots \\
a_2+b_2 w+\ldots\end{pmatrix},
$$
then using $SL(2)$ gauge freedom one can shift
\begin{equation}\label{Hecke:gaugeone}
(b_1,b_2)\mapsto (b_1,b_2)+c (a_1,a_2),\quad c\in\CC
\end{equation}
as well as rescale
\begin{equation}\label{Hecke:gaugetwo}
(a_1,a_2,b_1,b_2)\mapsto \lambda(a_1,a_2,b_1,b_2),\quad
\lambda\in\CC^*
\end{equation}
without changing $E_+$. The transformation (\ref{Hecke:gaugeone})
can be gotten rid of by replacing $(b_1,b_2)$ with an invariant
combination $y=b_1a_2-b_2 a_1$. Then the space of Hecke
transformations is parameterized by a triple $(a_1,a_2,y)$ modulo a
rescaling
$$
(a_1,a_2,y)\mapsto (\lambda a_1,\lambda a_2,\lambda^2
y),\quad\lambda\in\CC^*.
$$
In addition, we have the condition that $(a_1,a_2)\neq (0,0)$
(because $s_1(0)\neq 0$). The resulting moduli space is the total
space of the bundle $\cO(2)$ over the projective line $\PP^1$ with
the homogeneous coordinates $(a_1:a_2)$.

Suppose now that the subspace $V$ is one-dimensional; then we get a
restriction on the choices of $s_1$ and $s_2$ by requiring the
holomorphic section of $E_-$ taking value in $V$ at $w=0$ to be a
linear combination of $s_1'$ and $s_2'$ with holomorphic
coefficients. This restriction says that at $w=0$ $s_1$ should also
take value in $V$. This fixes the parameters $(a_1,a_2)$ up to an
overall scaling, and the resulting space of allowed Hecke
modifications is isomorphic to the fiber of $T\PP^1$ over the point
$(a_1:a_2)$.

Finally, if the subspace $V$ is two-dimensional (which is a generic
situation), then the space of allowed Hecke transformations is
empty. This means that applying an 't Hooft operator to a generic
$(G,R)$ Higgs bundle (regarded as a 0-brane on $\MV$) at a generic
point $p\in\Sigma$ gives the zero object in the category of branes.

\section{S-duality and an $\cN=2$ version of the Geometric Langlands
Duality}\label{Sduality}

It is conjectured that some or maybe all finite $\cN=2$
super-Yang-Mills theories enjoy S-duality which exchanges electric
and magnetic charges, Wilson and 't Hooft operators, and strong and
weak coupling. S-duality is best understood in the case when $R$ is
the adjoint representation and $G$ is arbitrary \cite{Osborn}, in
which case one has $\cN=4$ supersymmetry, or in the case when
$G=SU(2)$ and $R$ is a sum of four copies of the fundamental
representation. In this section we discuss in turn these two cases.

\subsection{Montonen-Olive conjecture and classical geometric
Langlands duality}

If $R$ is the adjoint of $G$, the theory has $\cN=4$ supersymmetry.
The Montonen-Olive conjecture states that $\cN=4$ theories with
gauge groups $G$ and $\LG$ are equivalent provided the gauge
couplings are related by $$\Ltau=-\frac{1}{n_\g \tau},$$ where
$n_\g$ is $1,2,$ or $3$ depending on the maximal multiplicity of
edges in the Dynkin diagram of the Lie algebra $\g$ of $G$
\cite{Doreyetal, AKS}. Langlands duality exchanges weight and
coweight lattices of $G$, and this is compatible with the exchange
of Wilson and 't Hooft operators under S-duality.

Let us draw the implications of the Montonen-Olive duality for the
$\beta''$-model regarded as an effective sigma-model on $C$. If $R$
is the adjoint representation, the nonabelian vortex equations on
$\Sigma$ become the usual Hitchin equations, and the target space of
the effective sigma-model on $C$ is $\MH(G,\Sigma)$. This moduli
space is hyperkahler, rather than K\"ahler, because of $\cN=4$
supersymmetry. The $\beta''$-model is equivalent to the B-model with
target $\MH$ taken with a complex structure in which $\delta A_\bw$
and $\delta q_w$ are holomorphic differentials. In this complex
structure $\MH$ is isomorphic to the moduli space
$\cM_{Higgs}(G,\Sigma)$ of semistable Higgs bundles; in \cite{KW}
this complex structure was denoted $I$.

The Montonen-Olive conjecture implies that the B-models for
$\cM_{Higgs}(G,\Sigma)$ and $\cM_{Higgs}(\LG,\Sigma)$ are
isomorphic. This implies that their categories of B-branes (i.e.
derived categories of coherent sheaves) are equivalent. More
specifically, both moduli spaces are fibered over an affine space
(this is known as the Hitchin fibration \cite{Hitchin}), and there
is an identification of the two affine spaces such that generic
fibers for $G$ and $\LG$ over the same point are dual abelian
varieties. Then if we exclude singular fibers from consideration,
the equivalence between the categories of B-branes is given by the
fiberwise Fourier-Mukai transform. The duality of Hitchin fibrations
for $G$ and $\LG$ is known as classical geometric Langlands duality
and has been proved for nonsingular fibers in
\cite{Faltings,HT,DP,HitchinG2}.

The classical geometric Langlands duality involves only
``commutative objects'' (coherent sheaves) on both sides. It is a
limit of the usual geometric Langlands duality in the following
sense. As explained in \cite{KW}, the usual geometric Langlands
duality is a consequence of mirror symmetry which relates the
B-model of $\MH(\LG,\Sigma)$ in complex structure $J$ and the
A-model of $\MH(G,\Sigma)$ in complex structure $K$. One can
continuously deform $J$ into $I$, therefore deforming the B-model in
complex structure $J$ into the B-model in complex structure $I$. On
the mirror side, this corresponds to deforming the symplectic
structure $\omega_K$ into a generalized complex structure, so that
in the limit one gets ordinary complex structure $I$. That is, on
the mirror side the A-model in complex structure $K$ is deformed
into the B-model in complex structure $I$.

\subsection{Seiberg-Witten duality}

Now let us turn to the case $G=SU(2)$, and $R$ being the sum four
copies of the fundamental representation. This theory is believed
\cite{SW2} to be self-dual under
$$
\tau\mapsto -\frac{1}{4\tau}.
$$
Together with a more obvious symmetry $\tau\mapsto\tau+\frac12$ this
generates $PGL(2,\ZZ)$ duality group.\footnote{The theory is
invariant under $\theta\mapsto \theta+\pi$ because amplitudes with
odd instanton number vanish because of fermionic zero modes.} The
action on $\tau$ takes a more familiar form if we rescale $\tau$ by
a factor $2$; then $\tau$ is acted upon by the group $PGL(2,\ZZ)$ of
integral fractional linear transformations in a standard way:
\begin{equation}\label{PGLtwo}
\tau\mapsto \frac{a\tau+b}{c\tau+d},\quad \begin{pmatrix} a & b \\ c
& d \end{pmatrix}\in SL(2,\ZZ).
\end{equation}
We will call this conjectural duality the Seiberg-Witten duality.
Although it is superficially similar to the Montonen-Olive duality,
there are also important differences. First of all, the SW duality
does not exchange $G$ and $\LG$. Second, on the Coulomb branch it
maps the 't Hooft-Polyakov monopole to a massive quark state, while
the massive W-boson is mapped to a bound state of two monopoles.
This suggests that the dual of the Wilson operator in the
fundamental representation will be a topologically trivial 't Hooft
operator with $SU(2)$ coweight corresponding to the adjoint
representation of $\LG=SU(2)/\ZZ_2$. As for topologically nontrivial
't Hooft operators carrying nonvanishing Stiefel-Whitney class $w_2$
(called the 't Hooft flux in the physics literature), they are not
allowed in the Seiberg-Witten theory because there are fields in the
fundamental representation.

Third, the SW duality acts nontrivially on the global ``flavor''
symmetry currents. For massless hypermultiplets, the theory has
$Spin(8)$ symmetry under which four copies of $q$ and four copies of
$\tq$ transform as an eight-dimensional vector representation. The
subgroup of $Spin(4)$ which does not mix $q$ and $\tq$ is isomorphic
to $SU(4)\times U(1)_B$. The generator
$$
S=\begin{pmatrix} 0 & 1 \\ -1 & 0\end{pmatrix}
$$
of the S-duality group acts on the $Spin(8)$ currents by an outer
automorphism known as triality. Triality maps the vector
representation of $Spin(8)$ to the the spinor representation. This
is compatible with the fact that quarks and charge-one monopoles are
vectors and spinors of $Spin(8)$, respectively \cite{SW2}.

The $\beta''$-twist made use of the $U(1)_B$ current. Thus in order
to extract the implications of the SW duality for the
$\beta''$-model, we must only consider the subgroup of the S-duality
group which preserves the $U(1)_B$ current. This subgroup consists
of elements of $PGL(2,\ZZ)$ congruent to the identity matrix modulo
$2$. It is denoted $\Gamma(2)$ in the mathematical literature and is
generated by
$$
T^2=\begin{pmatrix} 1 & 2 \\ 0 & 1\end{pmatrix}
$$
and
$$
S T^2 S=\begin{pmatrix} -1 & 0 \\ 2 & -1\end{pmatrix}
$$
Thus the Seiberg-Witten duality conjecture predicts that $\Gamma(2)$
acts by autoequivalences on the derived category of $\MV$.

It is easy understand the autoequivalence corresponding to the
element $T^2$. The transformation $T^2$ is simply a shift of
theta-angle by $2\pi$. In the $\beta''$-model, it shifts the B-field
by the 2-form
$$
-\frac{i}{2\pi}\int_\Sigma d^2w\,\Tr\ \delta A_\bw\wedge \delta
A_w=\frac{1}{4\pi}\int_\Sigma\Tr\ \delta A\wedge\delta A.
$$
This 2-form is minus the pull-back of the curvature of the
determinant line bundle $\Det$ on $\Bun(SL(2),\Sigma)$ via the
obvious projection $\MV(SU(2),R,\Sigma)\ra \Bun(SL(2),\Sigma)$.
Since the 2-form is of type $(1,1)$, it has no effect on the
$\beta''$-model if $C$ has no boundaries. In the presence of
boundaries, such a shift can be undone by tensoring the B-brane by
the the pull-back of $\Det$. Thus $T^2$ corresponds to tensoring all
B-branes on $\MV$ by the pull-back of $\Det$.

On the other hand, the transformation $ST^2S$ is not visible at weak
coupling, and its effect on a general B-brane is harder to describe.
An additional constraint on it comes from considering the action of
Wilson-'t Hooft operators on branes. Wilson operators have a very
simple effect on 0-branes on $\MV$: a 0-brane corresponds to a
particular $(G,R)$ Higgs bundle $(E,\frq)$ on $\Sigma$, and a Wilson
operator in representation $\cS$ inserted at a point $p\in\Sigma$
has the effect of tensoring the 0-brane with the vector space
$\cS(E)_p$. Thus any 0-brane $(E,\frq)$ is an eigenobject for a
Wilson operator inserted at any point $p\in\Sigma$, and all the
``eigenvalues'' fit into a holomorphic vector bundle $\cS(E)$. The
autoequivalence $T^2$ obviously commutes with all Wilson operators,
but $ST^2S$ maps Wilson operators to mixed Wilson-'t Hooft
operators. Thus the image of a 0-brane under $ST^2S$ must be an
eigenobject for certain Wilson-'t Hooft operators.

More explicitly, for $SU(2)$ we can identify both the lattice of
weights and the lattice of coweights with integers, and the Weyl
group $\cW\in\ZZ_2$ acts by negation. Thus a general loop operator
is described by a pair $(m,n)\in\ZZ^2$ modulo an overall sign
change, where $m$ and $n$ are the coweight and weight, respectively.
A Wilson loop with the same electric charge as that of a quark
corresponds to the pair $(0,1)$, while the 't Hooft-Polyakov
monopole has the same magnetic charge as the 't Hooft operator
$(1,0)$. More generally, a Wilson loop in $n+1$-dimensional
irreducible representation of $SU(2)$ corresponds to the pair
$(0,n)$. Under the transformation (\ref{PGLtwo}) the pair $(m,n)$
transforms as follows:
$$
\begin{pmatrix} m & n\end{pmatrix}\mapsto \begin{pmatrix} m & n\end{pmatrix}\begin{pmatrix} d & -b \\
-c & a\end{pmatrix}
$$
The pair $(0,n)$ is mapped by $ST^2S$ to $(2n,-n)$. The image of a
0-brane under the autoequivalence $ST^2S$ must be an eigenobject of
such Wilson-'t Hooft operators for all $n$ and all $p\in\Sigma$. To
understand what this means concretely, one needs to know how general
Wilson-'t Hooft operators act on B-branes; we plan to address this
important issue elsewhere \cite{future}.

In the case of geometric Langlands duality an important role is
played by the Hitchin fibration \cite{Hitchin}. As mentioned above
(see also \cite{KW} for a more detailed discussion), the Hitchin
fibration is a holomorphic map
$$
\cM_{Higgs}(G,\Sigma)\ra \cV,
$$
where $\cV$ is a complex affine space of dimension $(g-1) \dim G$.
This map is a holomorphic completely integrable system, in the sense
that its generic fibers are complex tori, and there is a canonical
holomorphic symplectic form on $\cM_{Higgs}(G,\Sigma)$ with respect
to which the fibers are Lagrangian. S-duality in this case acts by
an involution on the base $\cV$ and by T-duality on the fibers. For
simply-laced groups, the involution is trivial.

Some aspects of this construction do generalize to the case of
general $R$.\footnote{We are grateful to Nigel Hitchin for
explaining this to us.} The analog of the Hitchin fibration is a map
from $\MV$ to the space of gauge-invariant polynomials in the Higgs
field $q_w$. In the case we are considering, the Higgs field is
really a collection of four holomorphic 1-forms $q^i_w dw,
i=1,\ldots,4,$ with values in a holomorphic $SL(2)$ bundle $E$ on
$\Sigma$. The gauge-invariant polynomials
$$
B^{ij}=q_w^i q_w^j
$$
form a skew-symmetric $4\times 4$ matrix with values in
$H^0(K_\Sigma^2)$. The entries of the matrix are not independent but
satisfy a constraint
$$
\eps_{ijkl} B^{ij} B^{kl}=0.
$$
Since $h^0(K^4_\Sigma)=7(g-1)$, we get $7(g-1)$ equations which
define a cone $\cC$ in the $18(g-1)$-dimensional affine space
$H^0(K_\Sigma^2)\otimes \CC^6$. This cone is the analog of the base
of the Hitchin fibration. It has the same dimension as $\MV$ and the
projection from $\MV$ to $\cC$ is generically one-to-one. That is,
a generic pair $(E,\frq)$ is determined by its image in $\cC$.

The full SW duality group acts nontrivially on the fields $\tB,B,M$
whose expectation values parameterize the Higgs branch of the
theory. This follows from the fact that it does not commute with
flavor $Spin(8)$ symmetry which acts nontrivially on the Higgs
branch. But the subgroup $\Gamma(2)$ that we are considering acts
trivially on the base of the Hitchin fibration (because both $T^2$
and $S^2=1$ act trivially). This is again similar to the situation
in the $\cN=4$ case.

\section{Concluding remarks}

In this paper we showed that any finite $\cN=2$ gauge theory can be
twisted into a field theory on $C\times\Sigma$ which is topological
along $C$ and holomorphic along $\Sigma$. The observables in the
twisted theory are independent of the gauge coupling and theta-angle
and are holomorphic functions of coordinates on $\Sigma$. One can
regard the twisted theory as providing a new class of ``protected''
observables which goes beyond the usual chiral ring.

We also showed that the twisted theory admits Wilson-'t Hooft loop
observables labelled by elements of the set
$\left(\Lambda_{cw}\times\Lambda_w\right)/\cW$. BRST symmetry forces
the loops to be of the form $\gamma\times p$, where $\gamma$ is a
loop in $C$ and $p\in\Sigma$. The OPE of two parallel loop operators
is nonsingular, therefore one gets a commutative algebra of
Wilson-'t Hooft operators. For purely electric (Wilson) loop
operators, this algebra is identical to the fusion algebra of
irreducible representations of $G$. For purely magnetic ('t Hooft)
operators, the algebra depends on the matter content of the theory.
In the case when the matter fields transform in the adjoint
representation of $G$, it follows from the results of \cite{KW} that
the algebra of 't Hooft operators is isomorphic to the fusion
algebra of irreducible representations of $\LG$. It remains an
interesting problem to describe the algebra of general Wilson-'t
Hooft operators. In the case when the matter is in the adjoint of
$G$, it should unify the representation theory of $G$ and $\LG$.
Perhaps it has something to do with the double-affine Hecke algebra
introduced by I.~Cherednik \cite{DAHA}.

We also proposed an analog of geometric Langlands duality for finite
$\cN=2$ gauge theories. It states that the S-duality group (or
rather, the subgroup of the S-duality group which commutes with the
twist) acts by autoequivalences on the derived category of the
moduli space of nonabelian vortex equations. This action should be
compatible with the action of Wilson-'t Hooft operators on the
derived category. In many respects, the $\cN=2$ and $\cN=4$
dualities are similar, but there are also important differences.
Perhaps the most important one is that unlike the Hitchin moduli
space, the moduli space of nonabelian vortex equations is not
holomorphic symplectic (although it is a Calabi-Yau), and
consequently the connection with algebraic integrable systems is
lost. An analog of the Hitchin map does exist, but its generic
fiber is not a complex torus; for example, for $G=SU(2)$ and
hypermultiplets in the fundamental representation the map is generically one-to-one.

\section*{Acknowledgments}

I would like to thank D.~Arinkin, R.~Bezrukavnikov, I.~Mirkovic, and
N.~Hitchin for explanations and discussions, and S.~Gukov,
N.~Hitchin, and N.~Saulina for comments on the preliminary draft of
the paper. I am also grateful to Ketan Vyas for performing some
computations related to the discussion of OPE of Wilson and 't Hooft
operators in section \ref{loopops}. This work was supported in part
by the DOE grant DE-FG03-92-ER40701.

\end{document}